\documentclass[fleqn,usenatbib,twocolumn]{mnras}

\usepackage[T1]{fontenc}
\usepackage{ae,aecompl}

\usepackage{graphicx}	
\usepackage{amsmath}	
\usepackage{amssymb}	
\usepackage{hyperref}
\usepackage[utf8]{inputenc}
\usepackage{mathrsfs}
\usepackage{bm}
\usepackage{hyperref}
\usepackage[dvipsnames]{xcolor}

\usepackage{soul}

\newcommand{\be}{\begin{equation}}	
\newcommand{\ee}{\end{equation}}	
	
\newcommand{\ms}{\,\mathrm{ms\ p.b.}}
\newcommand{\refig}[1]{Fig.~\ref{#1}}
\newcommand{\refeq}[1]{Eq.~\ref{#1}}

\newcommand{\LTWI}{low $T/|W|$ instability}

\newcommand{\hydro}{H}
\newcommand{\aligneddip}{L1-0}
\newcommand{\tilteddip}{L1-90}
\newcommand{\quadA}{L2-0A}
\newcommand{\quadB}{L2-0B}

\title[MM signatures from 3D MHD CCSN]{Three-dimensional core-collapse supernovae with complex magnetic structures: II. Rotational instabilities and multi-messenger signatures}

\author[M. Bugli et al.]{
M Bugli,$^{1}$\thanks{E-mail: matteo.bugli@cea.fr}
J. Guilet,$^{1}$
T. Foglizzo,$^{1}$
and M. Obergaulinger$^{2,3}$
\\
$^{1}$Laboratoire AIM, CEA/DRF-CNRS-Universit\'e Paris Diderot, IRFU/D\'epartement d'Astrophysique, CEA-Saclay F-91191, France\\
$^{2}$Departamento de Astronomía y Astrof\'isica, Universitat de Val\`encia, Dr. Moliner 50, 46100, Burjassot, Spain\\
$^{3}$Institut fur Kernphysik, Theoriezentrum, Schlossgartenstr. 2, D-64289 Darmstadt, Germany
}

\date{Accepted XXX. Received YYY; in original form ZZZ}

\pubyear{2022}

\usepackage{newtxtext,newtxmath}

\begin{document}
\label{firstpage}
\pagerange{\pageref{firstpage}--\pageref{lastpage}}
\maketitle

\begin{abstract}
The gravitational collapse of rapidly rotating massive stars can lead to the onset of the low $T/|W|$ instability within the central proto-neutron star (PNS), which leaves strong signatures in both the gravitational wave (GW) and neutrino emission.
Strong large-scale magnetic fields are usually invoked to explain outstanding stellar explosions of rapidly rotating progenitors, but their impact on the growth of such instability has not yet been cleared.    
We analyze a series of three-dimensional magnetohydrodynamic models to characterize the effects of different magnetic configurations on the development of the low $T/|W|$ and the related multi-messenger features.
In the absence of magnetic fields, we observe the growth on dynamical time scales of the low $T/|W|$, associated with a strong burst of GW and a correlated modulation of the neutrino emission.
However, models with a strong magnetic field show a quenching of the low $T/|W|$, due to a flattening of the rotation profile in the first $\sim100$ ms after shock formation caused by the magnetic transport of angular momentum.
The associated GW emission is weakened by an order of magnitude, exhibits a broader spectral shape, and has no dominant feature associated with the PNS large-scale oscillation modes.
Neutrino luminosities are damped along the equatorial plane due to a more oblate PNS, and the only clear modulation in the signal is due to SASI activity.
Finally, magnetized models produce lower luminosities for $\nu_e$ than for $\bar{\nu}_e$, which is connected to a higher concentration of neutron-rich material in the PNS surroundings.

\end{abstract}

\begin{keywords}
supernovae: general -- MHD -- instabilities -- gravitational waves -- neutrinos -- magnetars  
\end{keywords}



\section{Introduction}

The explosion of a massive star as a core-collapse supernova (CCSN) following its gravitational collapse is one of the most energetic events in the Universe, in which a huge amount of gravitational binding energy (typically $\sim10^{53}$ erg) is released mostly ($\sim99\%$) in the form of neutrino radiation.
The remaining fraction of the energy budget is enough to allow the onset of the explosion and energize an otherwise stalling shock-wave, which can be detected in the electromagnetic spectrum only after several hours from the gravitational collapse.  
The vast majority of CCSN are driven by the \emph{neutrino-heating mechanism}, which relies on the deposition of energy by a small fraction of the neutrinos emitted from the PNS into the hot material behind the stalling shock.
This exchange counters the energy losses due to the dissociation of Fe-group elements crossing the shock and the compression due to the ram pressure of the infalling layers of the star, hence pushing the shock outwards and leading to a supernova explosion \citep{janka2012}.
The explosion dynamics and the evolution of the central PNS during the first few hundred milliseconds after shock formation are crucial to determine the energetics of the supernova explosion and the properties of the compact object that will remain afterwards.
In particular, fluid instabilities  play a fundamental role in increasing the efficiency of the energy deposition by neutrinos.
The neutrino-driven post-shock convection is a prominent example, as it tends to increase the dwelling time of shocked material within the gain region \citep{foglizzo2006} and thus its absorption of neutrinos and anti-neutrinos.
Another important process whose onset can favor the launch of a CCSN is the \emph{Standing Accretion Shock Instability} \citep[SASI; ][]{blondin2003,foglizzo2007}, which boosts the neutrino-heating mechanism through the development of large-scale oscillations of the shock front.  

Rotation is another physical process that has a deep impact on the properties of the CCSN.
While in axisymmetric models the inclusion of rotation tends to reduce the neutrino luminosities and weaken the onset of the explosion, three-dimensional numerical models have shown that the onset of SASI non-axisymmetric spiral modes can compensate that effect and reinforce the shock's expansion \citep{kazeroni2017,summa2018}.  
Although most stellar progenitors of CCSN are expected to have mild to very slow rotation, the collapse of a massive stars with strong rotation can lead to the onset of the \emph{\LTWI{}} \citep{ott2005}, i.e. a type of rotational instability that produces large-scale non-axisymmetric modes in the PNS and its surroundings. 
This instability was first observed by \cite{shibata2003} in numerical models of isolated differentially rotating stars, with the fundamental mechanism responsible for its onset presented, most notably, by \cite{watts2004} and \cite{passamonti2015}.
While the \LTWI{} has also been detected in CCSN numerical simulations \citep{ott2005,takiwaki2016}, the impact of the PNS environment on the properties of the instability is not yet clear \citep{takiwaki2021}.
Nonetheless, it is clear that such process can be crucial for the explosion dynamics, as the forming spiral structures enhance the transport of energy within the gain region, thus promoting the shock expansion \citep{takiwaki2016,shibagaki2021}.

Unfortunately, no electromagnetic signal from the central core of the star can be detected at this stage of the explosion to infer the properties of the explosion mechanism, given the large photon optical thickness that characterises the interior of stars up to their photosphere.
However, one way to directly probe the dynamics of the explosion central engine is to detect the gravitational wave (GW) signal from the central PNS in the first few seconds of the gravitational collapse \citep{ott2009}.
Oscillation modes in the the dense and hot PNS are generally expected to be excited by convection (and in some cases by SASI), hence deforming the PNS on sufficiently short time-scale to produce a significant emission of GW \citep{andresen2017}.
Recent numerical models of fast rotating progenitor showing the onset of the \LTWI{} \citep{shibagaki2020,takiwaki2021,shibagaki2021} predict also a corresponding strong emission of GW that outshines all other physical sources of GW from the PNS and could be easily detected by the current generation of ground-based observatories for a galactic event.   

The other channel that provides a way to investigate the first instants of the gravitational collapse consists in the detection of (anti)neutrinos, which are continuously being produced in all three flavors until the end of the PNS cooling phase through a series of different reactions: beta-processes, thermal pair-production/annihilation, and scattering with nuclei, nucleons and leptons \citep{janka2017}.
The neutrino signal during the accretion phase (i.e. right after the neutronisation peak at shock formation) contains important information on the early explosion dynamics, as it can be modulated by SASI \citep{tamborra2014a}, display signatures of the \emph{Lepton-number Emission Self-sustained Asymmetry} \citep[LESA; ][]{tamborra2014,kuroda2020} and the \LTWI{} \citep{shibagaki2021}.
Overall, the GW and neutrino signals can display a high degree of correlation not only in their arrival time, but also in their spectral shape in the frequency domain, since they can be affected by the same physical process \citep{kuroda2017,shibagaki2021}.
This proves to be an important property of the multi-messenger emission from CCSN, as it can be exploited to optimize the search of GW signal in the data produced by ground-based observatories.  

Since most of the studies investigating the impact of the \LTWI{} on the explosion dynamics and the associated multi-messenger emission consider non-magnetized fast rotating progenitors \citep{takiwaki2016,shibagaki2020,takiwaki2021,shibagaki2021}, it remains unclear how the onset of the instability would be affected by the action of magnetic fields.
\cite{kuroda2020} reported an MHD numerical model that suggested the co-existence of the \LTWI{} and the kink instability, but there was no conclusive analysis that could clearly establish how the properties of the hydrodynamic instability would change in a magnetized progenitor.
This question becomes particularly important when we consider that in fast rotating progenitors the action of dynamo mechanisms within the PNS is highly favored, whether they are triggered by convection \citep{thompson1993,raynaud2020,raynaud2022,masada2022}, the magnetorotational instability \citep[MRI; ][]{akiyama2003,obergaulinger2009,reboul-salze2021,reboul-salze2022,guilet2022} or the Tayler instability \citep{barrere2022}.
In this work we analyse the series of 3D MHD models presented in \cite{bugli2021} to establish for the first time the impact of strong large-scale magnetic fields of different topologies on the onset of the \LTWI{} (Section 3.1) and its characteristic signatures on the GW and neutrino emission (respectively sections 3.2 and 3.3).
We will focus particularly on the multi-messenger emission that is most affected by the formation of non-axisymmetric structures, i.e. along the rotational axis for GW and in the equatorial plane for the neutrinos.


\section{Numerical setup}
The numerical models presented in this study were first presented in \cite{bugli2021} and were produced using the \texttt{AENUS-ALCAR} code \citep{obergaulinger2008,just2015}.
They all follow the gravitational collapse and subsequent explosion of the massive and rapidly rotating stellar progenitor \texttt{35OC} \citep{woosley2006}, which has $M_\mathrm{ZAMS}=35M_\odot$, sub-solar metallicity, a large iron core of mass $M_\mathrm{Fe}\approx2.1M_\odot$ and radius $R_\mathrm{Fe}\approx2.9$ km, and rotates at its centre with angular velocity $\Omega_c\approx10$ rad/s.
While we retained the original rotation profile provided by the stellar evolution model, we replaced its magnetic fields with ad-hoc profiles of different topologies.

Model \hydro{} is a hydrodynamic benchmark with no magnetic fields, which produces a delayed neutrino-driven explosion at $t\sim400$ ms p.b.
Among the simulations presented here, \hydro{} produces not only the weakest explosion ($\sim10^{50}$ erg), but also the most massive and rapidly rotating PNS, due to the lack of efficient magnetic extraction of rotational energy.

Model \aligneddip{} has instead a strong aligned dipolar magnetic field, which leads to a prompt magneto-rotational explosion displaying symmetric bipolar outflows and the most energetic ejecta among the simulations we performed ($\sim1.4\times10^{51}$ erg after $\sim$400 ms from shock formation).  
Its magnetic field is defined as the curl of a vector potential whose only component in the azimuthal direction is given by
\begin{equation}\label{eq:Aphi}
    A^\phi_l(r,\theta)= r\frac{B_0}{(2l+1)}\frac{r_0^{3}}{r^{3}+r_0^{3}} \frac{P_{l-1}(\cos\theta)-P_{l+1}(\cos\theta)}{\sin\theta}.
\end{equation}
In the previous expression the multipolar order $l$ is set to 1, while $B_0$ is a normalisation constant that corresponds to the strength of the magnetic field along the vertical axis divided by a factor $\sqrt{l/2\pi}$, $r_0$ is the radius of the region where the strength of the field is roughly constant and $P_l$ is the Legendre polynomial of order $l$.

Simulations \quadA{} and \quadB{} are initialized with an aligned quadrupolar magnetic field of similar intensity. Although they differ for the relativistic correction to the Newtonian gravitational potential they adopt \citep{marek2006}, they produce very similar results: a weaker magneto-rotational explosion with slower and less collimated ejecta.
The main difference with model \aligneddip{} is an enhanced outward transport of angular momentum close to the equatorial plane (due to a non-vanishing radial field), but a weaker one across the polar regions, thus leading to a less efficient magneto-rotational mechanism.  
Their magnetic field is also constructed using \refeq{eq:Aphi}, but setting $l=2$. 

Finally, model \tilteddip{} considers a progenitor with an equatorial dipole (i.e. tilted by $90^\circ$ w.r.t. to \aligneddip{}), a configuration justified by the results presented in \cite{reboul-salze2021,reboul-salze2022} which show highly tilted dipolar components in the magnetic field amplified by the MRI within the PNS.
In this case the vector potential is given by \cite{halevi2018}
\begin{align}
    A_r & =  0, \\
    A_\theta & =  -r\frac{B_0}{2}\frac{r_0^{3}}{r^{3}+r_0^{3}}\sin\phi,\\
    A_\phi & =  -r\frac{B_0}{2}\frac{r_0^{3}}{r^{3}+r_0^{3}}\cos\theta\cos\phi.
\end{align}
This simulation produces a delayed explosion with a rather spherical shock front (similarly to model \hydro{}) but showing at the same time a well collimated, high-entropy column along the rotational axis (as the magnetised models).

All models consider $r_0=1000$ km and $B_0=10^{12}$, which results in a total magnetic energy in the numerical grid of $E_\mathrm{mag}\approx10^{48}$ erg and a ratio to the total rotational energy $E_\mathrm{mag}/E_\mathrm{rot}\approx0.1$.
The numerical grid is resolved uniformly along the $\theta$ and $\phi$ directions respectively with 64 and 128 points.
Along the radial direction, instead, the grid has a uniform resolution of $\Delta r=0.5$ km up to $r\sim10$ km, where the aspect ratio of the grid cell's surface faces becomes roughly uniform, i.e. $\Delta r \approx r\Delta\theta$.
Beyond this radius, the radial grid is logarithmically stretched up to $r_\mathrm{max}\approx 8.8\times10^4$ km, for a total of 210 points.   
The details on the adopted equation of state \citep[LS220, ][]{lattimer1991} and the settings of the neutrino transport using an M1 method can be found in \cite{bugli2020,bugli2021}.

\begin{figure}
    \centering
    \includegraphics[width=0.50\textwidth]{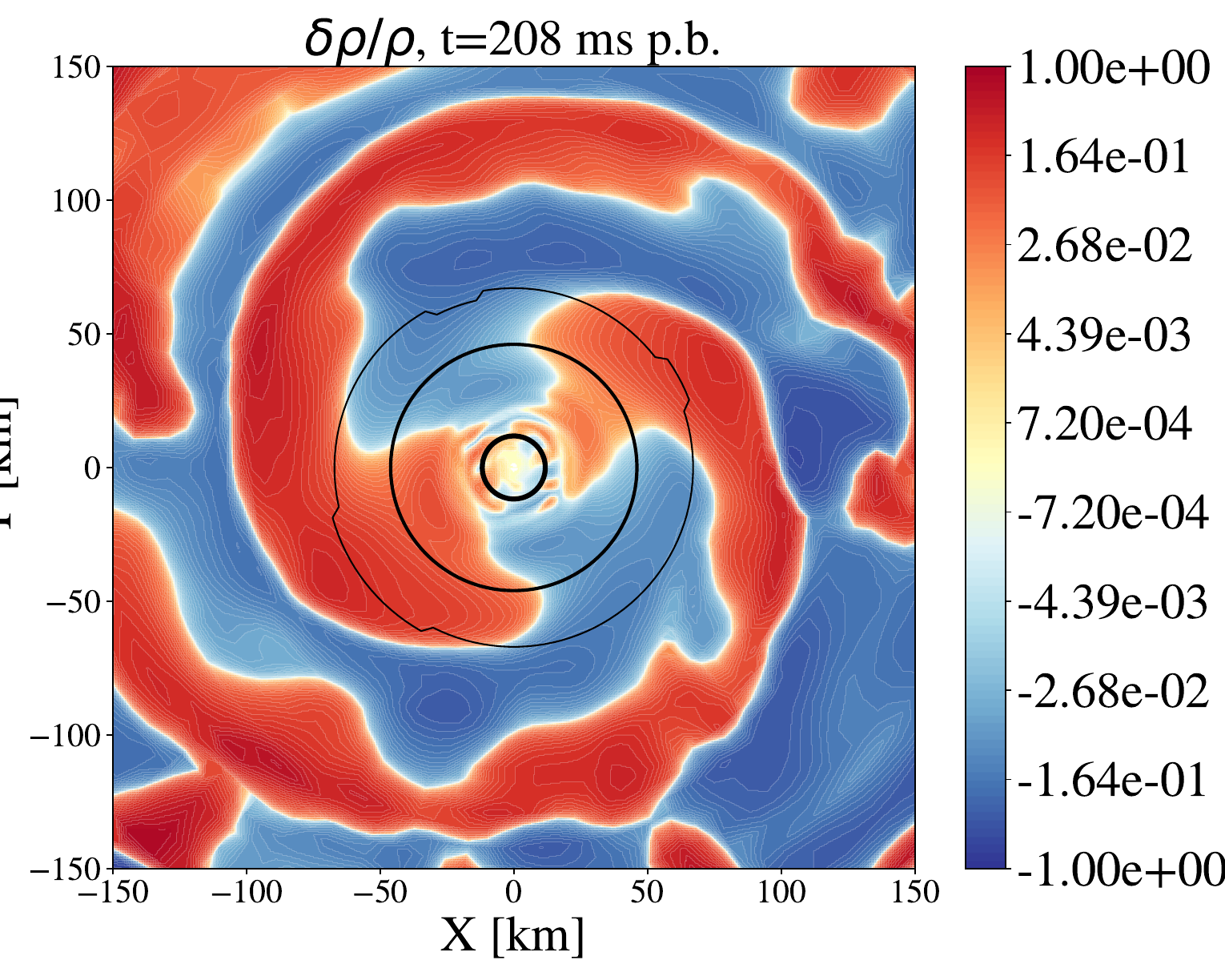}
    \includegraphics[width=0.50\textwidth]{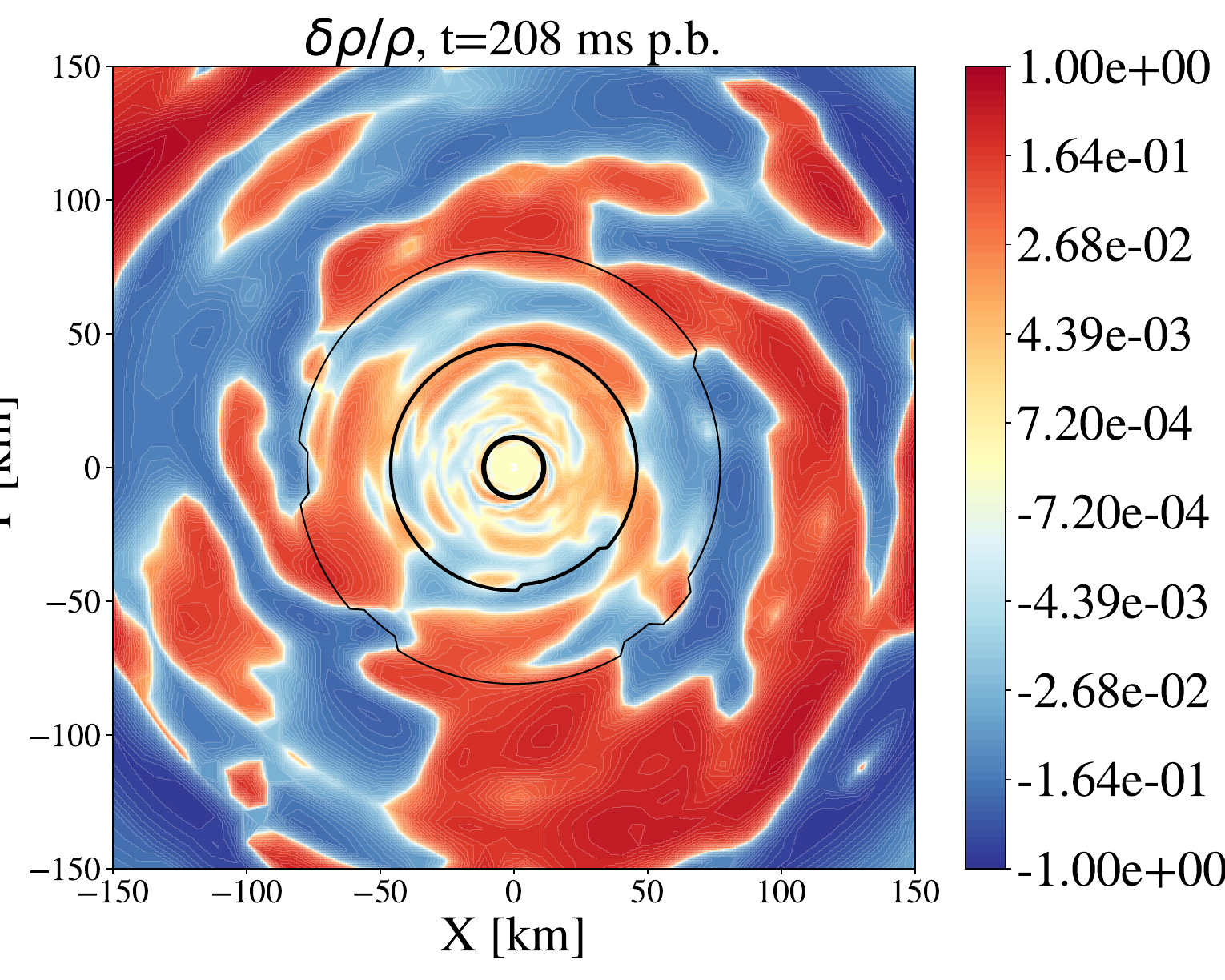}
    \caption{Relative deviations from the axisymmetric density distribution in the equatorial plane for models \hydro{} (top panel) and \aligneddip{} (bottom) at $t\sim208\ms$.}
    \label{fig:deltarho_equator}
\end{figure}

\section{Results and discussion}   


\subsection{The \LTWI{}}
Model \hydro{} is subject to a strong growth of the \LTWI{}, which clearly manifests itself at $t\sim200\ms$ through the formation of two-armed spiral structures in the density distribution across the equatorial plane (\refig{fig:deltarho_equator}).
The arms develop within the PNS and extend well beyond it throughout the post-shock region.
Their clear emergence at $t\sim200\ms$ marks the beginning of the non-linear saturated phase of the instability, which follows the linear growth of the \LTWI{} occurring in the first hundreds of ms after the core bounce.
By contrast, model \aligneddip{} presents density perturbations with no dominant azimuthal mode, having also lower amplitudes in the denser regions of the PNS.
A qualitatively similar scenario is offered by the other magnetized models, where no clear large-scale spiral structures develop during the post-bounce phase.

\begin{figure}
    \centering
    \includegraphics[width=0.5\textwidth]{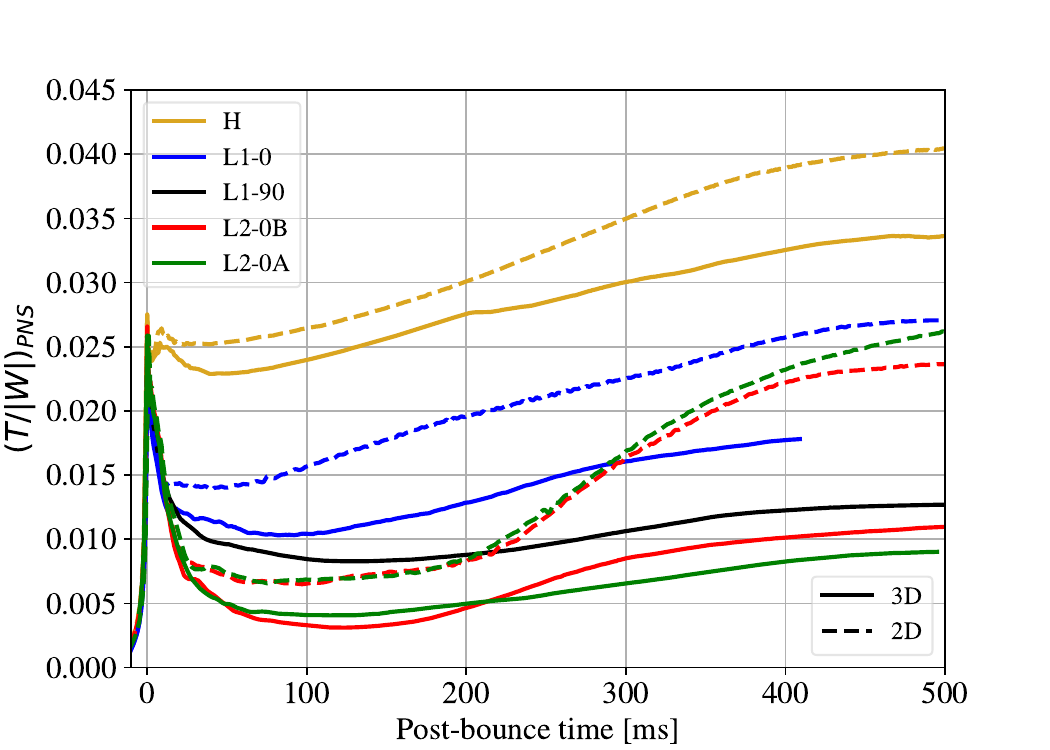}
    \caption{Times series of $T/|W|$ evaluated inside the PNS for all 2D (dashed lines) and 3D (solid lines) models presented in \protect\cite{bugli2021}.}
    \label{fig:T/W}
\end{figure}

We compute the ratio of rotational kinetic over gravitational energy within the PNS as
\be
\frac{T}{|W|}=\frac{\int\frac{1}{2}\rho v_\phi^2\mathrm{d}\bm{V}}{|\int\rho \Phi\mathrm{d}\bm{V}|},
\ee
where $\rho$ is the mass density, $v_\phi$ is the azimuthal velocity, $\Phi$ is the gravitational potential and we identify the volume of the PNS as the region where the matter density exceeds $10^{11}$ g/cm$^3$.
\refig{fig:T/W} confirms the different character of the hydrodynamic model compared to the others: the value of $T/|W|$ within the PNS volume is considerably larger for model \hydro{}, in particular due to the fact that it does not drop dramatically within the first few tens of ms after bounce like in the magnetized models.
However, there is still a transient decrease by $\sim10\%$ of $T/|W|$ until 50 ms p.b., after which a steady increase follows.
This feature in the evolution of $T/|W|$ has been seen in the numerical models presented in \cite{shibagaki2021} but it is completely absent in the simulations analysed in \cite{takiwaki2021}, which could be due to their use of the spherical approximation for the most central regions of the PNS. 
Models initialized with an aligned quadrupolar magnetic field show the strongest drop. Compared to them, \aligneddip{} keeps a value of $T/|W|$ which is systematically larger after $\sim20$ ms p.b. 
The simulation with an equatorial dipole displays instead a shallower decrease and increase of $T/|W|$, with values that roughly fall in between those for the aligned dipole model and simulations with a quadrupolar magnetic field.
Although axisymmetric models cannot capture the development of corotation instabilities, it is interesting to note how in the first $\sim15$ ms p.b. there is no remarkable difference with the corresponding 3d models.
At later times, instead, the value of $T/|W|$ becomes systematically larger in 2d, which is a direct consequence of the enhanced outward transport of angular momentum occurring in 3d simulations.

\begin{figure}
    \centering
    \includegraphics[width=0.5\textwidth]{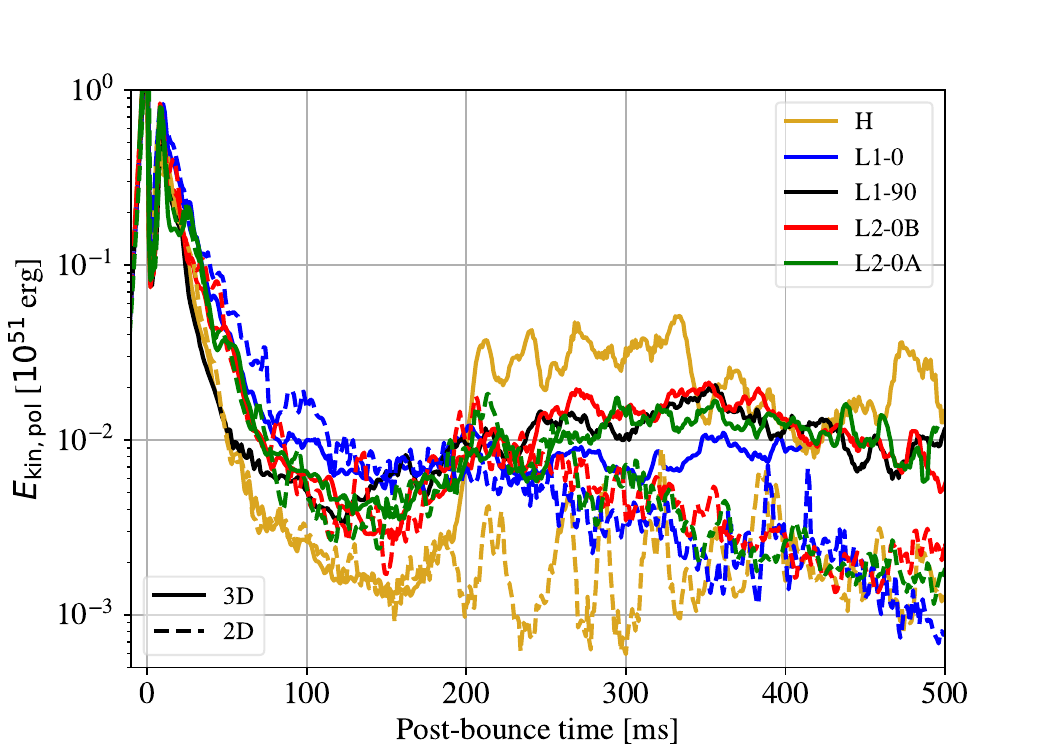}
    \caption{Time series of the poloidal kinetic energy in the PNS. }
    \label{fig:pns_energy}
\end{figure}

The behaviour of $T/|W|$ is inherently connected to the dynamical evolution of the PNS rotation profile and kinetic energy distribution.
In \refig{fig:pns_energy} we reported $E_\mathrm{kin,pol}$, i.e. the kinetic energy associated to the poloidal velocity field of the PNS for both 3d and 2d models.
The different problem's dimensionality has little effect on the evolution of $E_\mathrm{kin,pol}$ up until $\sim200$ ms p.b., where curves start deviating significantly from each other.
In all magnetized 3d models the poloidal kinetic energy ceases to decrease between 100 and 200 ms p.b., attaining a roughly constant level of $\sim10^{49}$ erg.
The non-magnetized simulation behaves in a qualitatively similar way, although in the first 200 ms $E_\mathrm{kin,pol}$ drops to a lower level, due to the lack of MHD-driven turbulence.
The poloidal kinetic energy then sharply increases by an order of magnitude within $\sim10$ ms, which coincides with the appearance of the spiral structures shown in \refig{fig:deltarho_equator}.

\begin{figure}
    \centering
    \includegraphics[width=0.5\textwidth]{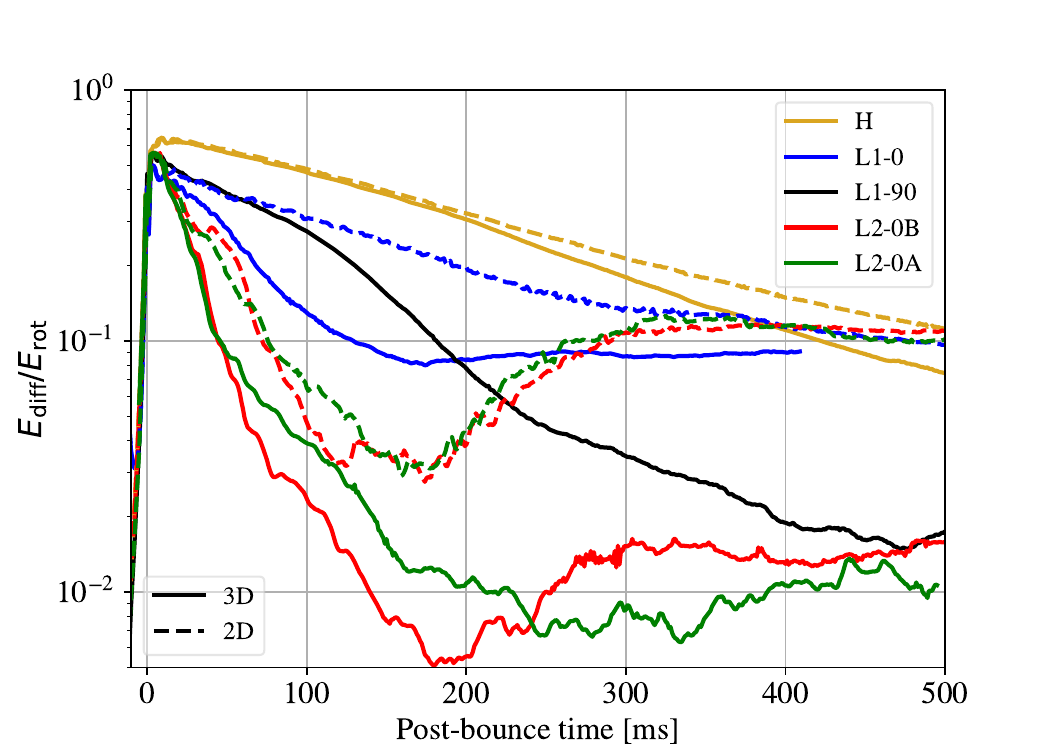}
    \caption{Fraction of rotational energy in the PNS due to differential rotation over time.}
    \label{fig:diff_rot}
\end{figure}

One of the fundamental ingredients required for the development of the \LTWI{} is not just a sufficiently fast rotation, but also a differential rotation profile \citep{watts2004,passamonti2015}.  
We convert the angular momentum, $L$, and the moment of inertia, $I$, of the PNS into the rotational energy of a rigidly rotating configuration and subtract it from the total rotational energy, $E_\mathrm{rot}$, to obtain the differential rotational energy as
\be
E_\mathrm{diff}=E_\mathrm{rot}-\frac{L^2}{2I},
\ee
which is displayed in \refig{fig:diff_rot}.  
At bounce, all models start from an approximate equipartition between rigid and differential rotation.  
The efficient outward transport of angular momentum in models with large-scale magnetic fields (in particular for the quadrupolar geometry with a non-vanishing radial field at the equator) causes the PNS and its surroundings to slow down within a few hundreds of ms.  
In the process, the rotational profile flattens and the degree of differential rotation decreases below 10\% (\aligneddip{}) or even 1\% (quadrupolar fields and tilted dipole).  
Except for the more gradual decline in model \tilteddip{}, the decrease occurs within the first 200 ms, after which the fraction of differential rotation levels off.  
The loss of differential rotation, while also potentially fast early after bounce, is less pronounced in the axisymmetric versions of these models and leads to a common value of $\sim10$\% of the overall rotational energy in the PNS.
Fig.~9 of \cite{bugli2021} exemplifies the key differences among our models in the angular momentum transport in the PNS, as it shows the loss rates of angular momentum through the PNS surface due to magnetic stresses in the equatorial and polar regions.
It should be noted that, in principle, angular momentum can also be advected away by the outflow itself, whose contribution can be simply estimated by replacing the factor $B_\phi\bm{B}_\mathrm{pol}$ with $\rho v_\phi\bm{v}_\mathrm{pol}$ in Eq.~9 of \cite{bugli2021}.
However, at any given time and for all our models there is almost never a net loss of angular momentum from the PNS through this process, but rather a gain due to the accretion of rotating matter.
The only exception is a transient loss of angular momentum in the first 50 ms p.b. in model \aligneddip{} connected to the prompt launch of the polar jet, which however subsides quickly later and is overall less than the loss due to magnetic stresses throughout the simulation.

These results clearly show how the action of magnetic fields deeply modifies the rotation profile of the PNS within the first 200 ms p.b., making the development of the \LTWI{} less likely in the process.
The hydrodynamic model experiences a slower decrease of its differential rotation with 10\% of its rotational energy still in the form of differential rotation at 500 ms p.b., thus retaining more favourable conditions for the \LTWI{} than the magnetized models.
It starts to significantly deviate from its axisymmetric version after around 200 ms p.b., when the loss of differential rotation of the 3d model accelerates.  
This deviation in fact coincides with the appearance of a genuinely non-axisymmetric feature in the form of the spiral structures shown in \refig{fig:deltarho_equator}, which indeed can contribute to the transport of angular momentum and the flattening of the rotation curve.

\begin{figure}
    \centering
    \includegraphics[width=0.5\textwidth]{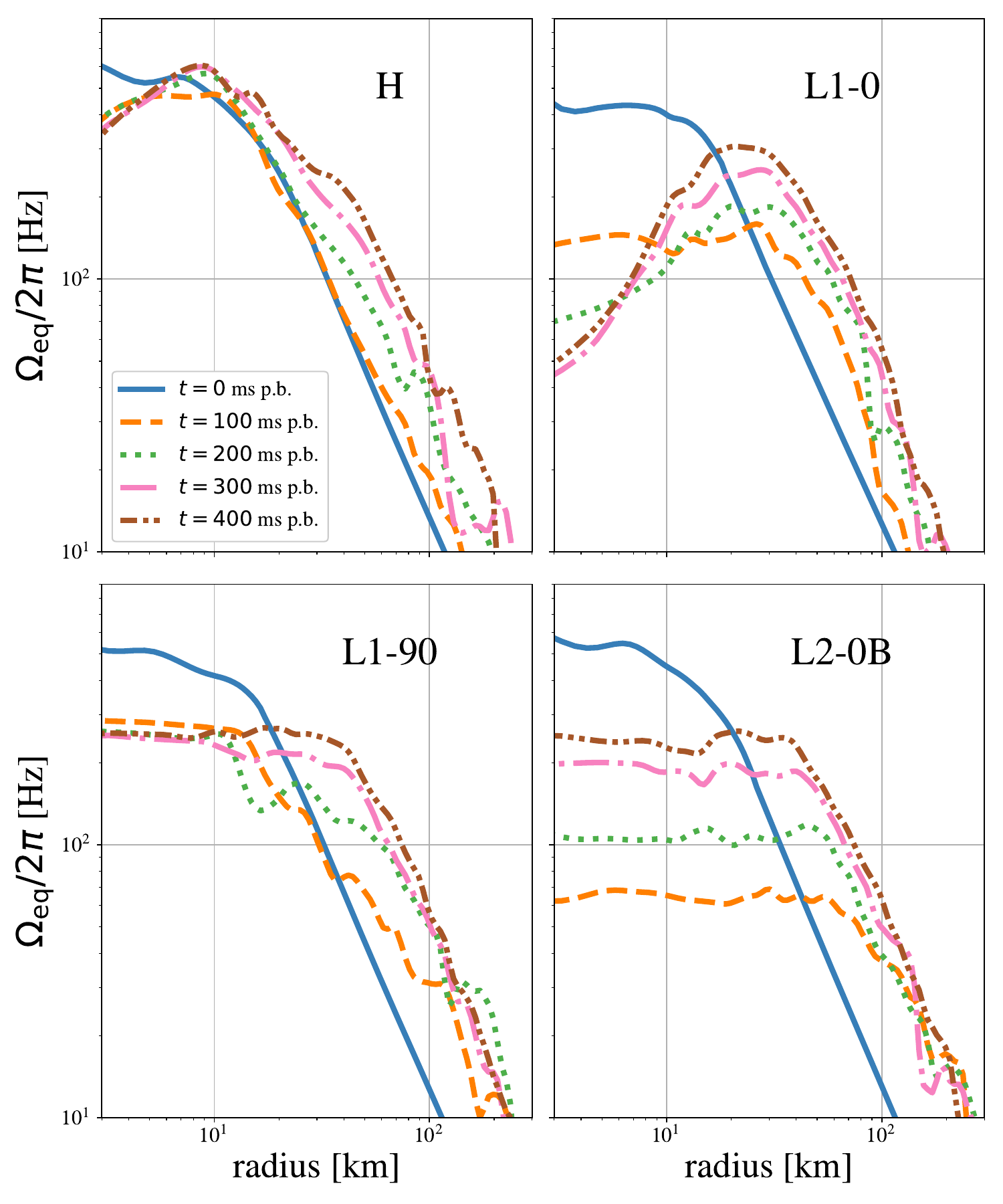}
    \caption{Equatorial radial profiles of the angular velocity $\Omega$ at a given azimuthal angle $\phi=0^\circ$ and different times for models \hydro{} (top left panel), \aligneddip{} (top right), \tilteddip{} (bottom left) and \quadB{} (bottom right).}
    \label{fig:omega_radius}
\end{figure}

The evolution of the equatorial radial profile of the angular velocity $\Omega_\mathrm{eq}$ is reported in \refig{fig:omega_radius}.
As expected, despite starting at bounce with a relatively similar rotation (blue curves), our 3d models soon evolve towards largely different profiles.
Model \hydro{}'s rotation has a fast transition towards a PNS with increasing angular velocity $\Omega$ in its inner core ($\sim10$ km), which then decreases with a powerlaw.
The slope of the decline in $\Omega$ at larger radii remains roughly constant over time, while the overall rotation increases steadily due to the contraction of the PNS and the accretion of material with high angular momentum.
The simulation with an aligned dipolar magnetic field offers a different scenario: after 100 ms, the inner regions of the PNS are in solid body rotation up to $\sim30$ km, and after that point the efficient extraction of angular momentum from the polar regions of the PNS slows down the rotation quite significantly.  
Model \tilteddip{} has a different evolution, since in the first 100 ms the inner core is also in solid body rotation, but with a higher velocity. 
Later on the central region remains in solid body rotation without slowing down, because the transport of angular momentum through the poles is much more inefficient w.r.t. the aligned dipole case.
However, the size of such region increases from 15 to 40 km, as the non-vanishing radial field at the equator favors the flattening of the rotation curve.
Finally, our simulation with an aligned quadrupolar magnetic field has the fastest transition (within the first 100 ms) towards a large inner region with a size of $\sim80$ km where the angular velocity is constant.
As pointed out in \cite{bugli2021}, this is due to the strong radial field with a coherent sign (as opposed to the equatorial dipole, where it reverts its polarity at multiple radii at the equator due to the winding induced by the differential rotation), which leads to a strong outer transport of angular momentum at low latitudes in the span of $\sim100$ ms.  
Afterwards, the profile of $\Omega$ is simply shifted towards higher values by the contraction and accretion onto the PNS, 
with the inner region in solid body rotation slightly shrinking in size and the outer regions conserving the same slope with radius. 

\begin{figure}
    \centering
    \includegraphics[width=0.5\textwidth]{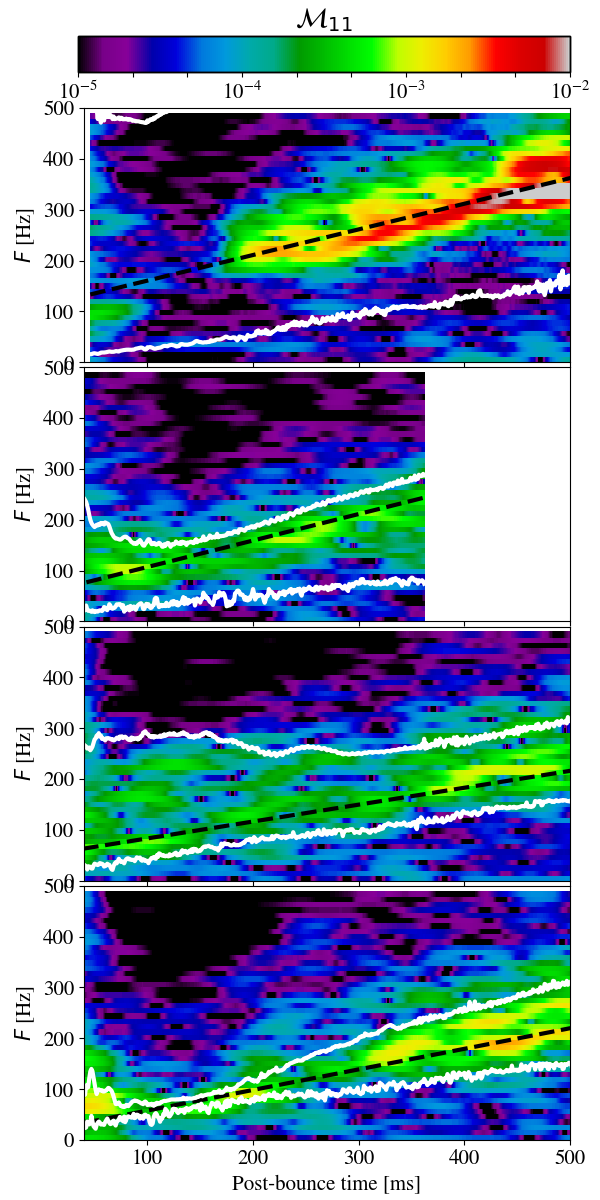}
    \caption{Time-frequency diagrams of $\tilde{\rho}_{11}/\tilde{\rho}_{00}$ for models \hydro{}, \aligneddip{}, \tilteddip{} and \quadB{} (from top to bottom).
    The black dashed line represents the linear fit of the mode, while the white curves show the rotation frequency at the PNS surface and the maximum rotation frequency within the PNS.}
    \label{fig:pns_rho_mode_11_spectrograms}
\end{figure}

\begin{figure}
    \centering
    \includegraphics[width=0.5\textwidth]{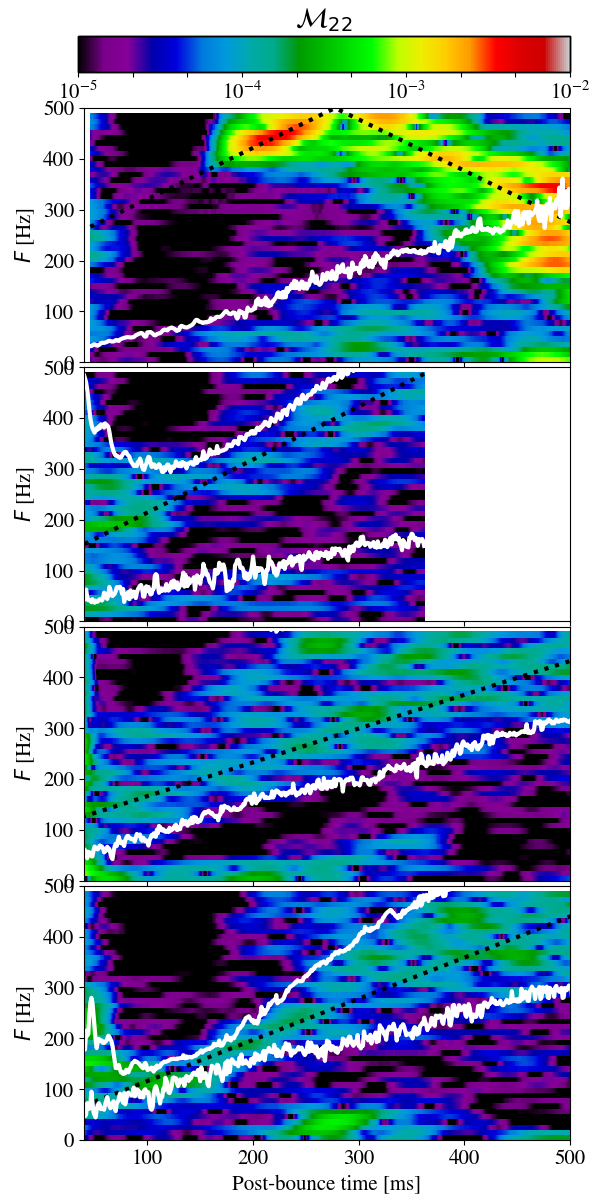}  
    \caption{Time-frequency diagrams of $\tilde{\rho}_{22}/\tilde{\rho}_{00}$ for models \hydro{}, \aligneddip{}, \tilteddip{} and \quadB{} (from top to bottom).
    The black dashed lines have twice the frequency of the dotted lines reported in \refig{fig:pns_rho_mode_11_spectrograms}. The white curves have been also multiplied by a factor $m=2$.}
    \label{fig:pns_rho_mode_22_spectrograms}
\end{figure}

We can characterize the development of the \LTWI{} more quantitatively by tracking the growth of its proper modes within the PNS.
To this end, we consider the decomposition in spherical harmonics of the density distribution over a sphere with radius $\tilde{r}=30$ km, i.e.

\be
\tilde{\rho}_{lm}(t)=\int \rho(r=\tilde{r},\theta,\phi) Y_{lm} \mathrm{d}\Omega,
\ee
where $Y_{lm}$ are the real spherical harmonics defined as
\begin{equation}
    Y_{lm}= \begin{cases}
                \sqrt{2}(-1)^m\mathrm{Im}[Y_l^{|m|}] &\quad\mathrm{if\ }m<0 \\
                Y_l^m &\quad\mathrm{if\ }m=0 \\
                \sqrt{2}(-1)^m\mathrm{Re}[Y_l^m] &\quad\mathrm{if\ }m>0,
            \end{cases}
\end{equation}
and 
\begin{equation}
    Y_l^m=\sqrt{\frac{2l+1}{4\pi}\frac{(l-m)!}{(l+m)!}}P_l^m(\cos\theta)e^{im\phi},
\end{equation}
are the complex spherical harmonics, with $P_l^m$ the Legendre polynomial of order $(l,m)$.
In \refig{fig:pns_rho_mode_11_spectrograms} we show the spectral evolution in frequency $F$ of $\tilde{\rho}_{11}/\tilde{\rho}_{00}$ over time, obtained by performing its short-time Fourier transform as
\be\label{eq:M11}
\mathcal{M}_{11}(t,F)=\int_{-\infty}^{\infty}\frac{\tilde{\rho}_{11}}{\tilde{\rho}_{00}}(t')H(t-\tau)e^{-2\pi iFt'}\mathrm{d}t',
\ee
where we set the Hann window function $H$ to have a width of 96 ms. 

Model \hydro{} (top-left panel) is characterized by a strong $(l=1,m=1)$ mode that emerges around $\sim200$ ms p.b. with an initial frequency of $\sim 200$ Hz, which gradually increases at a rate of 0.50 Hz/ms.
On the other hand, our magnetized models display a growth of $\mathcal{M}_{11}$ earlier after bounce, but with much smaller amplitudes and a lower oscillation frequency.
The frequency shift over time is similar to the one found in the hydrodynamic model for simulation \aligneddip{} (about $0.52$ Hz/ms, on the top-right panel), while for the models with an equatorial dipole and aligned quadrupolar field it is slower (0.33 and 0.40 Hz/ms, respectively).
In all cases the $(l=1,m=1)$ mode has at least one corotation point inside the PNS, since its frequency falls within the rotation frequency range set by its value at the PNS surface and its maximum value within it (white lines on \refig{fig:pns_rho_mode_11_spectrograms}).
As our output data was sampled every ms, our analysis is limited to a maximum frequency of 500 Hz.
This explains the disappearance of the white curve in the top left corner of the first panel in \refig{fig:pns_rho_mode_11_spectrograms}), since the maximal PNS rotation frequency in the hydrodynamic model exceeds that limiting value quickly after bounce.

We performed a similar analysis for the $(l=2,m=2)$ mode, which is directly connected to the two-arm spiral structures shown in \refig{fig:deltarho_equator}.
Once again, the corotation point lies within the PNS, since the mode's frequency is enclosed by the rotational frequencies shown also in \refig{fig:pns_rho_mode_11_spectrograms} multiplied by a factor $m=2$.
The mode emerges in model \hydro{} at $200$ ms p.b., with double the frequency of the corresponding $(l=1,m=1)$ mode (about $400$ Hz).
Up to $t\sim290$ ms p.b. $\mathcal{M}_{22}$ displays also a shift in frequency which is twice the one found for the lower order mode, but then it appears to gradually decrease for the following hundred of ms.
This is an artifact due to the aliasing of the signal produced in our simulations, which is once again due to the limited output sampling frequency.
Such analysis is confirmed by the good agreement between the mode's feature in the spectrogram and a mirrored slope after the maximum frequency is reached (top panel of \refig{fig:pns_rho_mode_22_spectrograms}), allowing us to assume that with a higher sampling frequency we would have observed a continuous increase in frequency of $\mathcal{M}_{22}$ after 290 ms.
The magnetized models also show such $(l=2,m=2)$ mode, though at a much weaker amplitude than both their $(l=1,m=1)$ mode and $\mathcal{M}_{22}$ from model \hydro{}.
Simulation \aligneddip{} presents broad-band features up until $t\sim150$ ms p.b., after which the mode's frequency is better defined.
The decrease of the maximum rotational frequency within the PNS (represented in \refig{fig:pns_rho_mode_22_spectrograms} with the higher white curve in the second panel from the top) is caused by the efficient magnetic braking of the PNS core, which in turn constrains the feature's spectral width from the top. 
The case of \tilteddip{} is similar at first, but the less efficient extraction of rotational energy sets a wider ranges of possible maximal frequencies within the PNS.
This leads to the spectral feature not being bounded at higher frequencies, but only from the bottom (i.e. by the faster rotation at the PNS surface).
Finally, model \quadB{} displays a very narrow frequency range within up to $t\sim200$ ms, which reflects a strong flattening of the PNS rotation curve.
The mode's spectral shape begins then to broaden as well, due to an increase in the maximum rotation rate caused by the continuous advection of angular momentum onto the PNS.
In all cases the shift in frequency corresponds rather well to double the shift measured for the $(l=1,m=1)$ mode (dashed black lines), which shows how the two modes are strongly coupled to each other.

It is interesting to note that, contrary to the results of \cite{shibagaki2020,shibagaki2021}, our model \hydro{} develops two strongly coupled modes with $m=1,2$ roughly at the same time, rather than producing a first transient emergence of the $m=1$ followed by a persistent $m=2$ mode.
This could be caused by the different rotation profile they used (an analytical distribution with cylindrical symmetry and constant specific angular momentum beyond 1000 km), the properties of the progenitors or a combination of the two effects.

\begin{figure}
    \centering
    \includegraphics[width=0.5\textwidth]{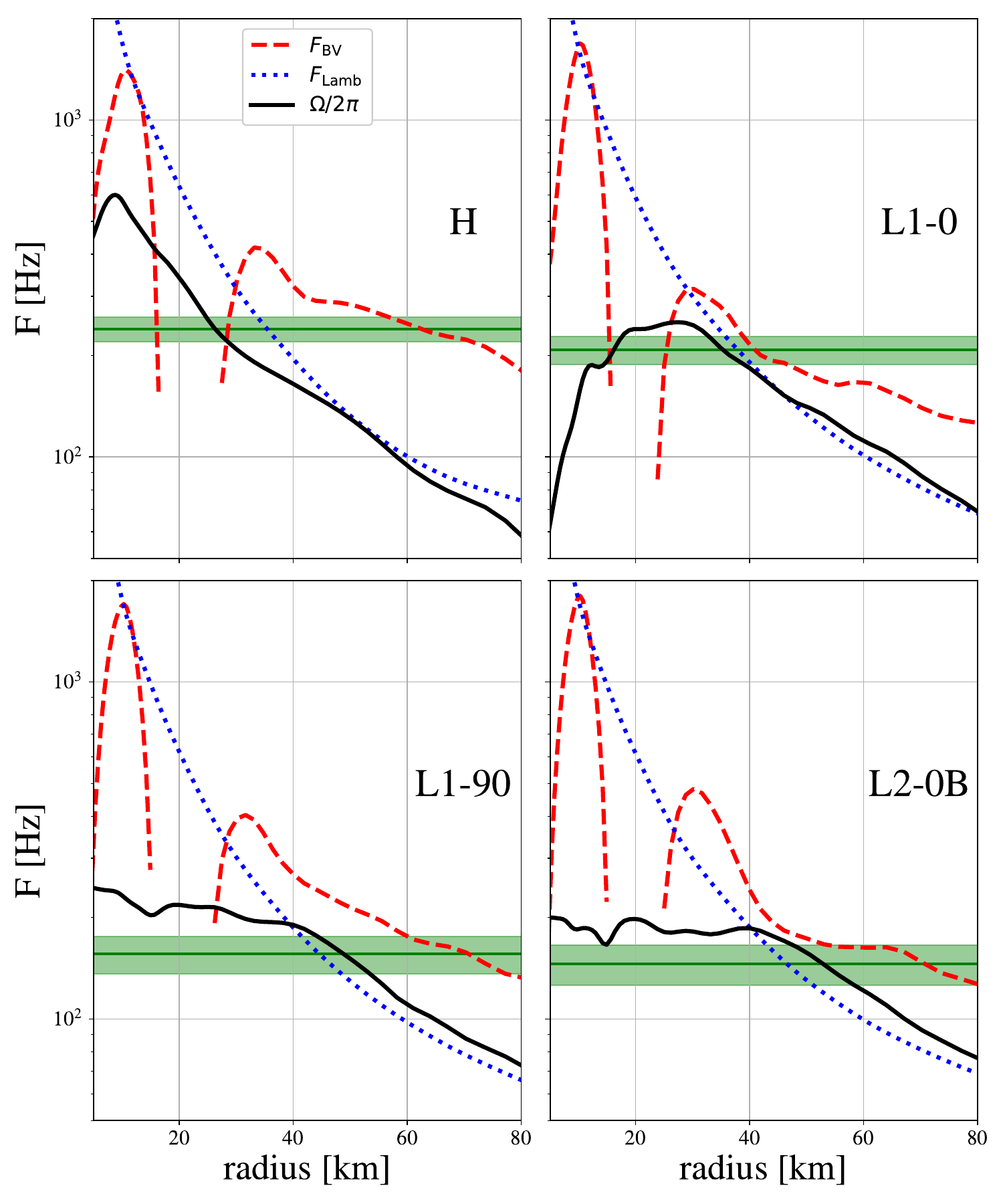}
    \caption{Radial equatorial  profiles of rotation (solid lines), Brunt-V\"ais\"al\"a (dashed) and  Lamb (dotted) frequencies at 300 ms p.b. for models \hydro{} (top left), \aligneddip{} (top right), \tilteddip{} (bottom left) and \quadA{} (bottom right).
    The green band represents the frequency of the $(l=1,m=1)$ mode estimated in \refig{fig:pns_rho_mode_11_spectrograms}.}
    \label{fig:freq_vs_radius}
\end{figure}

Although all our models show the growth of non-axisymmetric modes inside the PNS, only simulation \hydro{} develops strong large-scale structures clearly connected to the onset of the \LTWI{}.
However, all magnetized models display large-scale non-axisymmetric oscillating modes (although with much lower amplitude than the case of model \hydro{}) whose frequency falls within the range covered by the PNS differential rotation.
The fact that such modes appear very early in the simulations (as opposed to the 200 ms p.b. delay for the \LTWI{} modes) suggests that the physical mechanism responsible for their excitation involves magnetic fields.
A possible explanation could be the development of the kink instability \citep{mosta2014a,obergaulinger2021,barrere2022} which affects all our magnetized models at a very early stage of the explosion \citep{bugli2021}.
In presence of rotation the modes of the kink instability are in fact expected to co-move with the fluid \citep{bodo2016,bodo2019}, which is consistent with the frequency range of the PNS modes displayed in \refig{fig:pns_rho_mode_11_spectrograms} and \refig{fig:pns_rho_mode_22_spectrograms}.
Another possible scenario could be the excitation of oscillations by the azimuthal MRI \citep{heinemann2009,kitchatinov2010}, whose fastest growing mode is marginally resolved in our simulations.
However, it is less clear how such mechanism might produce in the CCSN context an $m=1$ deformation with a high degree of coherence in the polar direction ($l=1$) as we observe in our simulations, since the azimuthal MRI tends to favor the formation of structures with very high vertical wave numbers \citep{foglizzo1995,ogilvie1996,papaloizou1997}. 
Finally, we cannot exclude a priori that in the magnetized case these oscillations are also due to the onset of some form of the \LTWI{} \citep{muhlberger2014}, given the current uncertainties in the models and our current understanding of the magnetic effects on this co-rotational instability.

Overall, our magnetized models show a very different evolution of the rotation profile over time w.r.t. the non-magnetized case.
While its precise evolution may also vary from one magnetized models to another (see \refig{fig:omega_radius}), a common feature is that the action of magnetic fields considerably flattens the radial rotation profile in the PNS.
In order to understand the dichotomy between magnetized and hydrodynamic models, it is useful to identify more quantitatively the location of the mode's corotation point.
\refig{fig:freq_vs_radius} shows the radial profile at $t=300$ ms of the rotational frequency, the Brunt-V\"ais\"al\"a frequency
\be
F_\mathrm{BV}=\frac{1}{2\pi}\left[
                                  \frac{1}{\rho}\frac{\partial\Phi}{\partial r}
                                  \left(
                                        \frac{1}{c_s^2}\frac{\partial p}{\partial r}-\frac{\partial\rho}{\partial r}
                                  \right)
                            \right]^{1/2}
\ee
and the Lamb frequency
\be
F_\mathrm{Lamb}=\frac{1}{2\pi}\frac{\sqrt(l(l+1))c_s}{r}
\ee
evaluated for $l=1$.
For model \hydro{} the corotation point (i.e. the intersection between $\Omega$ and the green band representing the $(l=1,m=1)$ mode's frequency) falls at the outer edge of the convective zone, i.e. where $F_\mathrm{BV}^2$ is negative.
This result is consistent with the findings of \cite{takiwaki2021}, who identified the important role of convection in creating favourable conditions for the development of the \LTWI{}.
On the other hand, in the magnetized case the convective zone is generally in solid-body rotation due to the early magnetic transport of angular momentum, which could explain the lack of strong non-axisymmetric structures.
Moreover, the corotation of the same mode in the magnetized models appears to be located farther out in the PNS, where the differential rotation has not been damped by the action of magnetic stresses and the rotational frequency is significantly lower.
It is also interesting to note that only in the hydrodynamic case the mode's corotation resides in a zone where the mode's frequency is lower than $F_\mathrm{Lamb}$, i.e. in a region where p-modes cannot propagate radially.


\subsection{Gravitational waves}

\begin{figure}
    \centering
    \includegraphics[width=0.50\textwidth]{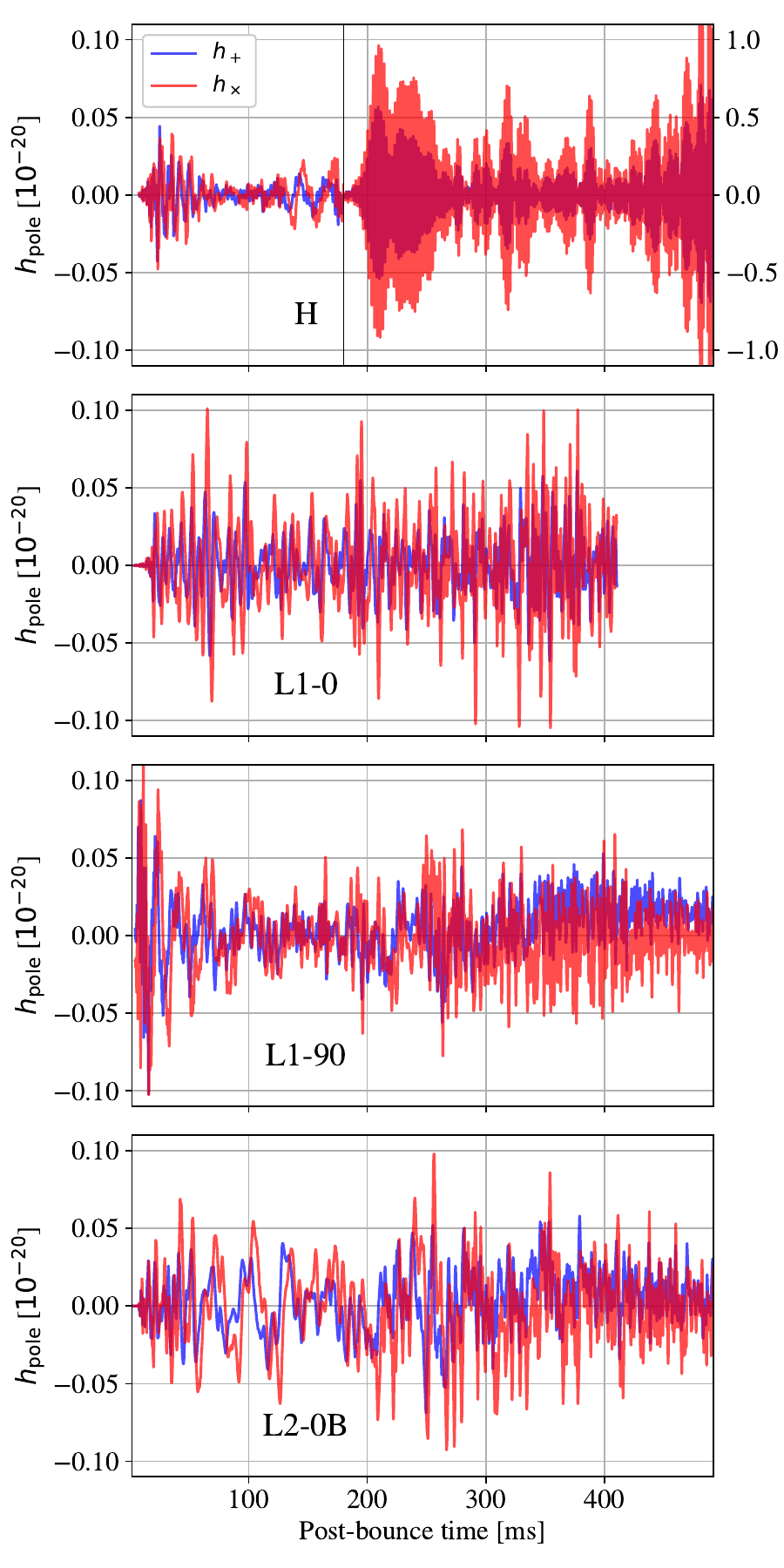}
    \caption{Gravitational wave strain $h$ measured along the northern axis assuming a distance of 10 kpc. 
    Blue and red curves represent the $+$ and $\times$ polarization components, respectively.
    The right portion of the first panel starting at $t=180$ ms p.b. (model \hydro{}) is rescaled by a factor 10 in the y-axis to capture the vastly increased GW signal.}
    \label{fig:gw_strain}
\end{figure}

\begin{figure}
    \centering
    \includegraphics[width=0.5\textwidth]{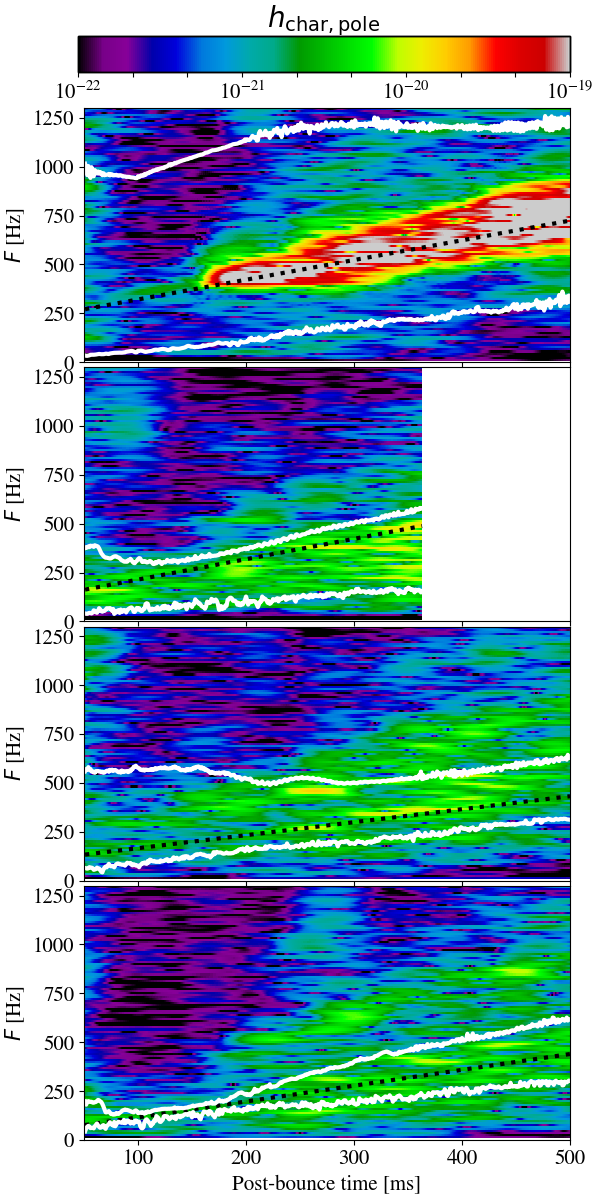}
    \caption{Time-frequency diagrams of $h_\mathrm{char}$ for models \hydro{}, \aligneddip{}, \tilteddip{} and \quadB{} (from top to bottom).
    The black dashed lines and the white curves have the same meaning as those drawn in \refig{fig:pns_rho_mode_22_spectrograms} (except for the lack of aliasing for the non-magnetized simulation).}
    \label{fig:gw_spectrogram}
\end{figure}

The development of non-axisymmetric perturbations within the PNS with multipolar order $l\geq2$ can be expected to leave a distinctive signature in the gravitational wave emission.

\refig{fig:gw_strain} shows the GW strain $h$ in its two polarizations (i.e. $+$ and $\times$) along the northern axis assuming a distance of $10$ kpc.
Model \hydro{} displays an initial emission of GWs with amplitude $\sim0.02$ which lasts for about 80 ms, followed by a brief quieter phase up until $\sim$125 ms p.b. and then lower frequency oscillations with amplitudes comparable to the initial emission.
This stage corresponds to the time at which a strong large-scale azimuthal modes are not observed in the PNS (\refig{fig:pns_rho_mode_11_spectrograms} and \refig{fig:pns_rho_mode_22_spectrograms}) and the turbulent kinetic energy experiences a steady decrease (\refig{fig:pns_energy}). 
At about $t\sim180$ ms p.b. the GW signal increases dramatically by two orders of magnitude, with the $\times$ polarization having a systematically higher amplitude w.r.t. the $+$ one.
The emission is not continuous, but is rather clustered into closely spaced multiple bursts of different amplitudes.
Such signal appears in correspondence with the increase in turbulent kinetic energy and the emergence of the non-axisymmetric modes within the PNS.
On the other hand, magnetized models show no such increase of the GW emission.
Simulations \aligneddip{} and \quadB{} initially present signals that are very similar to those produced in the hydrodynamic model.
However, the signal's amplitudes do not decrease as in the latter case, but rather remain of the order of $\sim0.05$ for the following hundreds of ms.
The model with an equatorial dipolar field is the only one showing a stronger initial emission in the first 30 ms p.b., which is due to the deformations induced by the intrinsically non-axisymmetric large-scale structure of the magnetic field .

The tight connection between the GW emission and the PNS modes becomes even more transparent when we consider the spectral evolution of the characteristic strain $h_\mathrm{char}$ over time, which is defined as \citep{kuroda2014}
\begin{equation}
    h_\mathrm{char}=\sqrt{\frac{2}{\pi^2}\frac{G}{c^3}\frac{1}{D^2}\frac{dE(\theta,\phi)}{dF}}.
\end{equation}
In the previous equation we introduced the GW spectral energy density
\begin{equation}
    \frac{dE(\theta,\phi)}{dF}=\frac{\pi}{4}\frac{c^3}{G}F^2\left(|\tilde{A}_+(F)|^2+|\tilde{A}_\times(F)|^2\right),
\end{equation}
with the Fourier components of the GW amplitude $A_{+/\times}(t)$ calculated as
\begin{equation}
    \tilde{A}_{+/\times}(F)=\int_{-\infty}^{\infty}A_{+/\times}(t)H(t-\tau)e^{-2\pi iFt}dt,
\end{equation}
similarly to the analysis of the PNS modes described by \refeq{eq:M11}.
The main difference from the time-frequency analysis conducted for the harmonic components in the PNS density distribution is the sampling frequency, which is 10 times higher for the GW signal.
This means that we can track spectral features in the GW emission up to 5 kHz, thus avoiding the aliasing effects observed at $\sim500$ Hz for $\mathcal{M}_{22}$ (first panel of \refig{fig:pns_rho_mode_22_spectrograms}). 

The first panel of \refig{fig:gw_spectrogram} exemplifies the connection between PNS modes and GW emission in model \hydro{} by revealing a strong feature around $t\sim180$ ms p.b.
The signal starts at a frequency of $\sim400$ Hz and ramps up in the diagrams at the same rate as the $(l=2,m2)$ mode shown in \refig{fig:pns_rho_mode_22_spectrograms} before the effects of aliasing become manifest.
The frequency of the bulk of the signal never exceeds the maximum rotational frequency in the PNS, which saturates after $t\sim250$ ms p.b. at about $1.2$ kHz.
On the other hand, all our magnetized models produce a GW emission with a broad-band spectral shape that is only initially contained within the boundaries set by the PNS rotation frequency for an $m=2$ mode, i.e $F^*=m\Omega/2\pi=\Omega/\pi$.
After 200 ms p.b. there is a significant emission of GW at frequencies $F>500$ Hz which cannot be associated to the \LTWI{} non-axisymmetric modes observed in the PNS but is most likely due to oscillation modes of the PNS similar to those generally observed in non-rotating CCSN numerical models \citep{sotani2016,torres-forne2019}.
The PNS modes shown in \refig{fig:pns_rho_mode_22_spectrograms} can still contribute to the emission at lower frequencies bounded by the rotation rate of the PNS (white curves in \refig{fig:gw_spectrogram}), but they cannot account for the whole signal produced in that band.
This results clearly show that the GW emission from models with dynamically strong magnetic fields is not dominated by the large-scale azimuthal modes induced by rotational instabilities.


\subsection{Neutrino signal}

\begin{figure}
    \centering
    \includegraphics[width=0.5\textwidth]{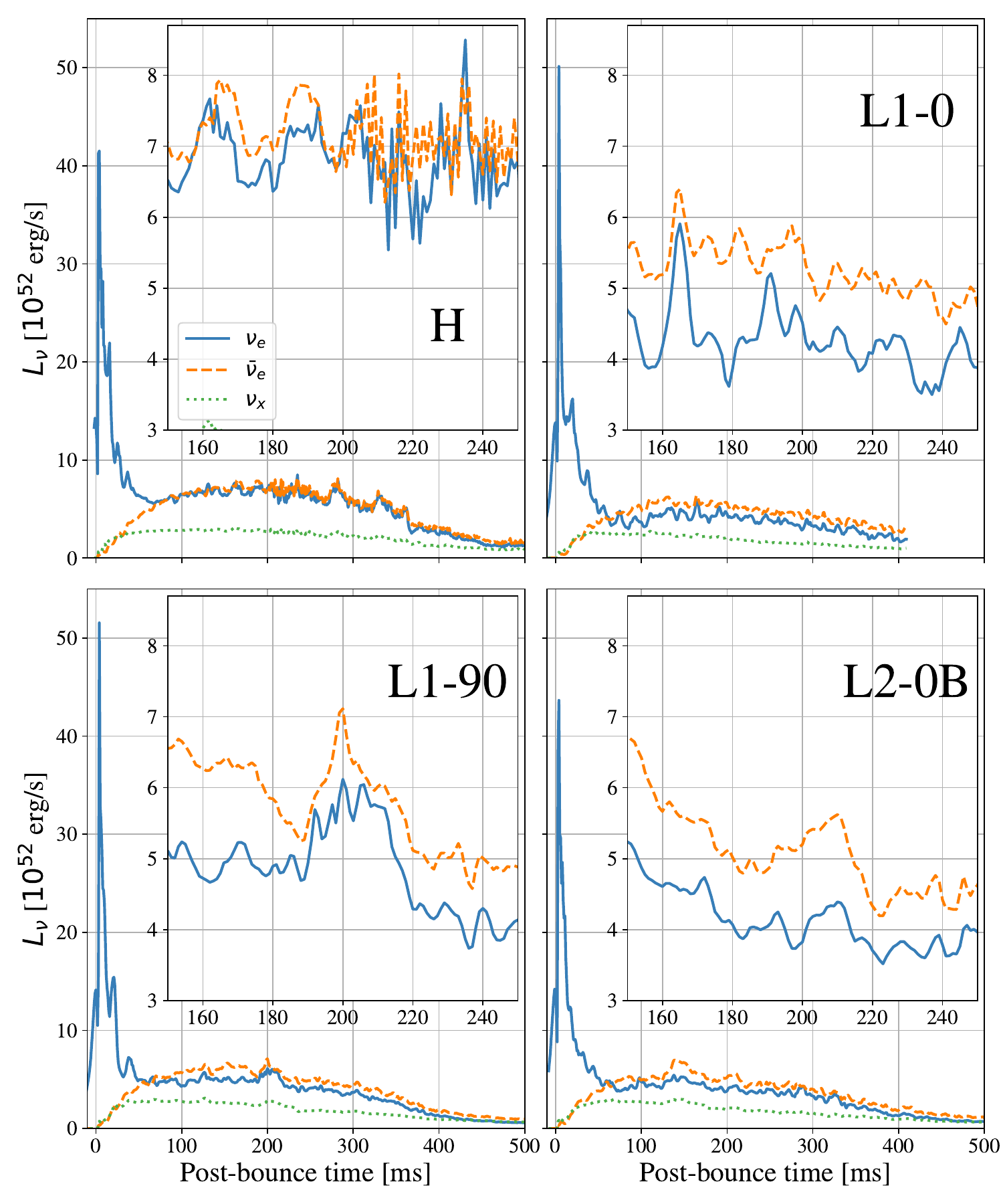}
    \caption{Equatorial neutrino lightcurves at a given azimuthal angle $\phi=0^\circ$ for all three species.}
    \label{fig:L_nu_timeseries}
\end{figure}

\begin{figure}
    \centering
    \includegraphics[width=0.5\textwidth]{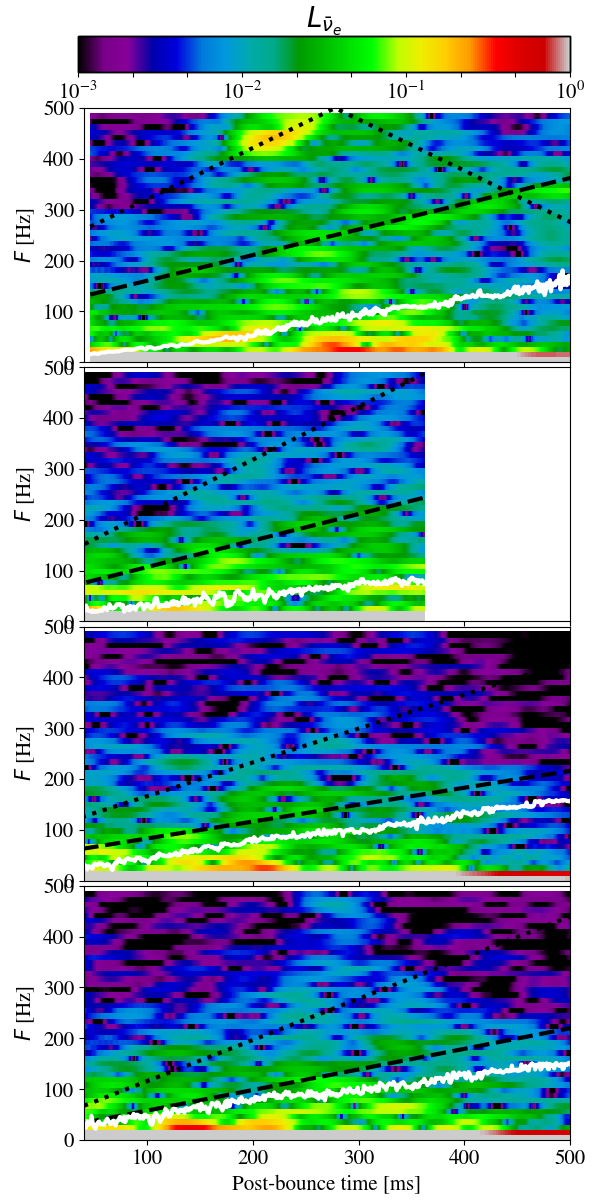}
    \caption{Time-frequency spectrograms of $\bar{\nu}_e$ lightcurves along the equatorial plane.
    The dashed and dotted curves represent the spectral patterns of the $(l=1,m=1)$ and $(l=2,m=2)$ modes in the PNS, while the white curve is the rotation frequency of the PNS surface.}
    \label{fig:L_nu_spectrograms}
\end{figure}

We now focus on the analysis of the neutrino emission from our 3d models.
\refig{fig:L_nu_timeseries} shows the luminosity curves for all three neutrino species at a distance of 500 km along the equatorial plane.
Once again we can recognize a strong dichotomy between the hydrodynamic model and the simulations that include magnetic fields.
Model \hydro{} is characterized by a higher luminosity for $\nu_e$ and $\bar{\nu}_e$ during the accretion phase w.r.t. the magnetized cases, while $L_{\nu_x}$ is relatively similar among the different models.
This is the result of the different accretion history onto the PNS in presence of magnetic fields, as the combination of more efficient outward transport of angular momentum and higher ejection of material through the polar outflows both contribute to decrease the neutrino luminosity.
Such an effect is not only reproduced among the axisymmetric versions of our models, but it is also consistent with the results of previous 2d simulations which considered magnetic fields with higher multipolar orders \citep{bugli2020}.
Another important feature of the neutrino emission from model \hydro{} is the appearance of fast oscillations after $t\sim200$ ms p.b. at a frequency of $\sim440$ Hz, which correspond to the onset of the large scale non-axisymmetric modes within the PNS.
By contrast, the magnetized models only show lower frequency modulations of the signal for the whole duration of the simulations.
Finally, in magnetized models the luminosity of $\nu_e$ is systematically lower than that of $\bar{\nu}_e$, while both species have very similar luminosities in absence of magnetic fields. 
Such difference can be observed as early as $60$ ms p.b., and lasts for the whole duration of the simulation.

The evolution of the spectral shape of the $\bar{\nu}_e$ signal shown in \refig{fig:L_nu_spectrograms} gives important insights on their physical origin.
In the top panel of \refig{fig:L_nu_spectrograms} we can recognize spectral features in the neutrino emission of model \hydro{} that correspond to the signatures of the non-axisymmetric PNS modes previously shown.
Contrary to GW (for which only perturbation with $l\geq2$ can produce a signal) one can expect deformations with $l=1$ to modulate the neutrino emission.
Indeed, the non-axisymmetric dipolar mode observed in \refig{fig:pns_rho_mode_11_spectrograms} appears in the $\bar{\nu}_e$ signal as well, with a clear pattern matching its frequency and time evolution.
The origin of the high-frequency oscillations can be identified in the emergence of the $(l=2,m=2)$ mode in the PNS, as a strong feature at roughly 440 Hz appears at $t\sim200$ ms p.b.
However, the footprint of its aliased component at later times appears to be weaker and less defined.
Contrary to the GW emission, the neutrino signal of model \hydro{} is not dominated by the non-axisymmetric modes within the PNS.
Most of the emission has a frequency lower than $\mathcal{M}_{11}$, with a burst between 250 and 400 ms p.b. 
The neutrino emission from the magnetized models, on the other hand, do not show a clear correlation with the non-axisymmetric modes of the PNS (especially w.r.t the quadrupolar modulation).
This suggests that only with a strong development of the \LTWI{} the resulting azimuthal modes can leave a clear signature on the neutrino signal. 
A similar broad-band, low-frequency emission is present in all cases, with isolated bursts at $F<50$ Hz for all models except \aligneddip{}.
A spectral analysis of $L_{\nu_e}$ and $L_{\nu_x}$ leads to very similar conclusions for all the models here considered.

\begin{figure}
    \centering
    \includegraphics[width=0.5\textwidth,clip,trim={0 1.25cm 0 0}]{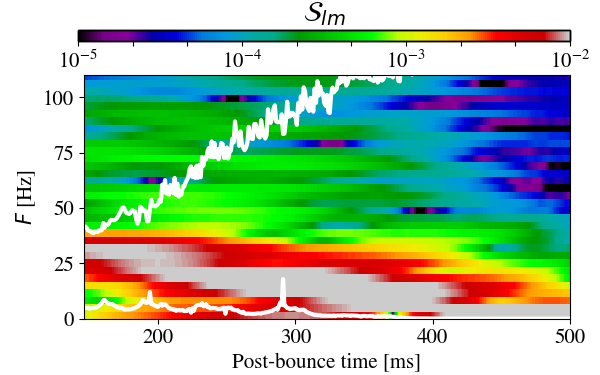}
    \includegraphics[width=0.5\textwidth,clip,trim={0 0 0 1.7cm}]{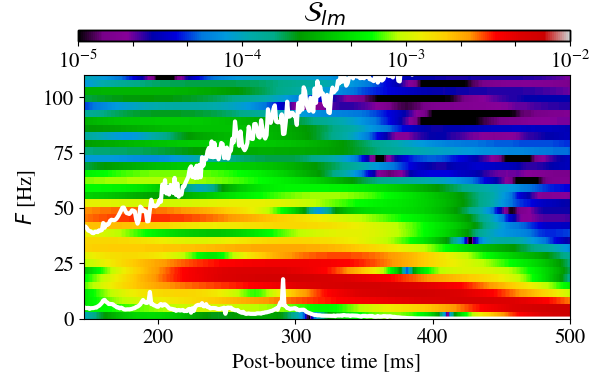}
    \caption{Time-frequency spectrograms of the shock radius non-axisymmetric modes $\mathcal{S}_{11}$ and $\mathcal{S}_{22}$.
    The white curves represent the rotational frequencies of the PNS surface (higher frequency) and the shock (lower) at the equator.}
    \label{fig:shock_spectrograms}
\end{figure}

\begin{figure}
    \centering
    \includegraphics[width=0.5\textwidth]{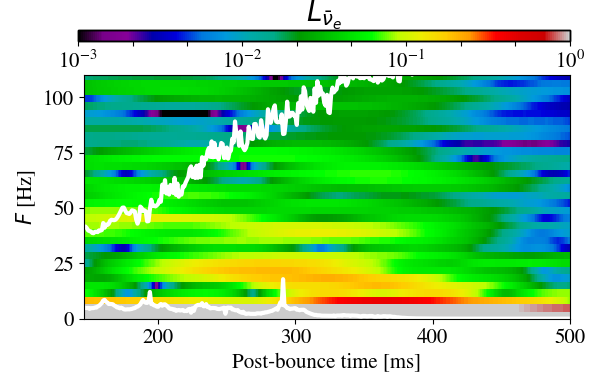}
    \caption{Same as the top panel of \refig{fig:L_nu_spectrograms} for model \hydro{}, but with increased Hann window function width $\tau=296$ ms.}
    \label{fig:L_nu_spectrograms_lowfreq}
\end{figure}

While the physical source of the low-frequency features of the neutrino signal cannot be the \LTWI{}, a mechanism such as SASI could instead provide a suitable explanation.
The development of SASI can indeed leave a clear signature in both the GW and neutrino emission in CCSN numerical models \citep{kuroda2014,kuroda2017,andresen2017,kuroda2020}. 
We performed for the shock radius $R_\mathrm{sh}$ a similar analysis to the one in section 3.1, computing its spherical harmonics components as
\be
\tilde{R}_{lm}(t)=\int R_\mathrm{sh}(\theta,\phi) Y_{lm} \mathrm{d}\Omega,
\ee
and then evaluating their spectral evolution with
\be\label{eq:S11}
\mathcal{S}_{lm}(t,F)=\int_{-\infty}^{\infty}\frac{\tilde{R}_{lm}}{\tilde{R}_{00}}(t')H(t-\tau)e^{-2\pi iFt'}\mathrm{d}t'.
\ee
The width of the Hann window function $\tau$ used in \refig{fig:L_nu_spectrograms} (96 ms) provides a good compromise between frequency and time resolution for the overall neutrino signal, but the resulting frequency resolution is too coarse to clearly appreciate slow spectral shifts at lower frequencies that characterize SASI.
In \refig{fig:shock_spectrograms} we show the time-frequency diagrams for the shock radius non-axisymmetric modes $\mathcal{S}_{11}$ and $\mathcal{S}_{22}$, which track the onset of SASI's spiral modes in model \hydro{}.
We set $\tau=296$ms to improve our frequency resolution $\Delta F$ from 10.4 Hz to 3.4 Hz, at the cost of increasing the spread over time of the spectral features.
The $l=1$ mode (top panel) displays a strong feature with a frequency of $\sim30$ Hz around $150$ ms p.b., which then slowly decreases in frequency over time down to the Hz scale.
The quadrupolar mode (bottom panel) shows a similar spectral shape, with a slightly higher early frequency of $\sim45$ Hz but a similar shift over time.
Other studies have found similar signatures in the neutrino emission \citep{tamborra2013, shibagaki2021}, although at a higher frequency ($\sim80$ Hz rather than $\sim30$ Hz).
This is likely due to our models presenting a generally larger advection time, which is connected to the initial thermodynamic profiles of the iron core and depends sensibly on the shock radius, the sound speed, and advection velocity in the post-shock region \cite{scheck2008}.
If we now compare these results with a higher frequency resolution of the spectrogram for $L_{\nu_e}$ (\refig{fig:L_nu_spectrograms_lowfreq}), we can clearly identify a very similar pattern consistent with those of the shock radius modes.
This shows how the low frequency non-axisymmetric SASI modes modulate the time variability of the neutrino signal in our hydrodynamic model.
We extended a similar analysis to the magnetized models, which has confirmed for simulations \tilteddip{} and \quadB{} the connection between the onset of SASI and the low frequency features in the neutrino emission signals in the last two panels of \refig{fig:L_nu_spectrograms}.
An exception to this trend is model \aligneddip{}, which contrary to the other simulations produces a prompt magnetorotational explosion.
This significantly quenches the development of SASI, thus leading to a weaker low-frequency modulation of the neutrino signal.

\begin{figure}
    \centering
    \includegraphics[width=0.5\textwidth]{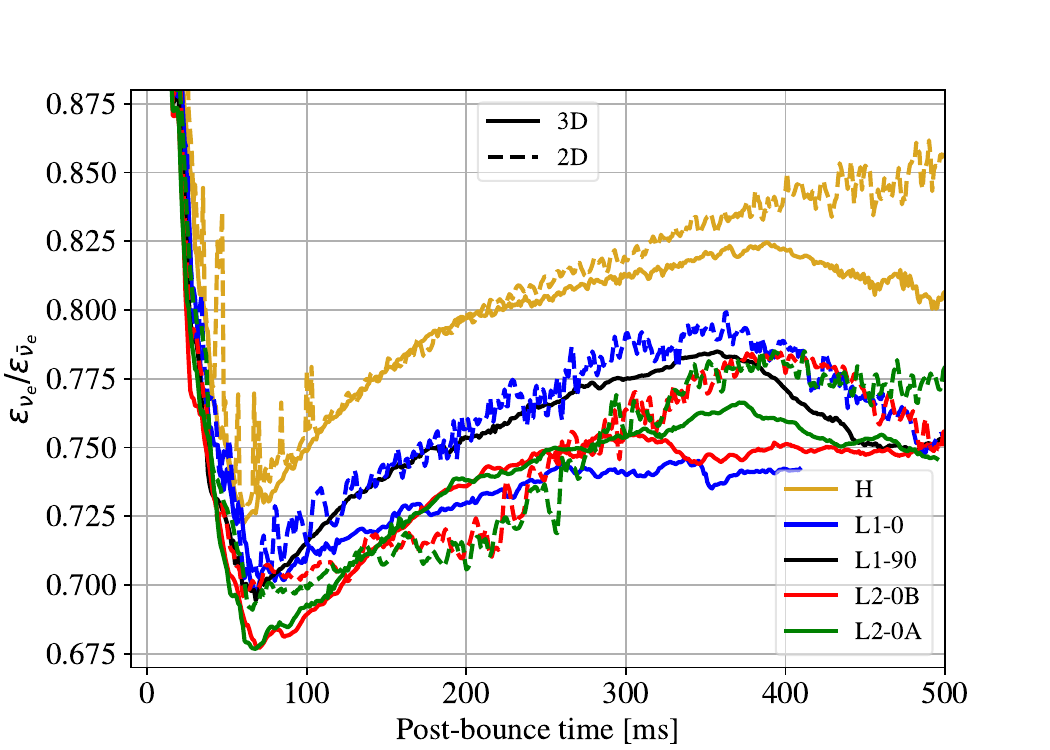}
    \caption{Ratios of electron neutrinos and anti-neutrinos mean energies over time at a distance of 500 km along the equator for different models.}
    \label{fig:nu_energies}
\end{figure}

To identify the cause of the difference between $L_{\nu_e}$ and $L_{\bar{\nu}_e}$ shown in \refig{fig:L_nu_timeseries} for the magnetized models we focus instead on the thermodynamic evolution of neutrino species.
\refig{fig:nu_energies} shows the ratio between the rms energies of $\nu_e$ and $\bar{\nu}_e$ at the same location considered for the lightcurves in \refig{fig:L_nu_timeseries}.
While model \hydro{} displays the highest value at any given time, the simulations with magnetic fields have a systematically lower ratio.
This points to the fact that the higher luminosities of electron anti-neutrinos are due to their higher energy at the emission site.
It is interesting to note that the same results hold for the corresponding axisymmetric models (dotted curves).
The hydrodynamic models shows deviations between the 2d and 3d realizations only after 200 ms p.b., which suggests a direct impact of the spiral modes caused by the \LTWI{} on the neutrino emission.
Finally, all 3d simulations tend to have a lower value of $\epsilon_{\nu_e}/\epsilon_{\bar{\nu}_e}$ w.r.t. their axisymmetric counterparts (in particular model \aligneddip{}), which is likely due once again to the more efficient transport of angular momentum in 3d.

\begin{figure}
    \centering
    \includegraphics[width=0.5\textwidth,clip,trim={0 2.8cm 0 0}]{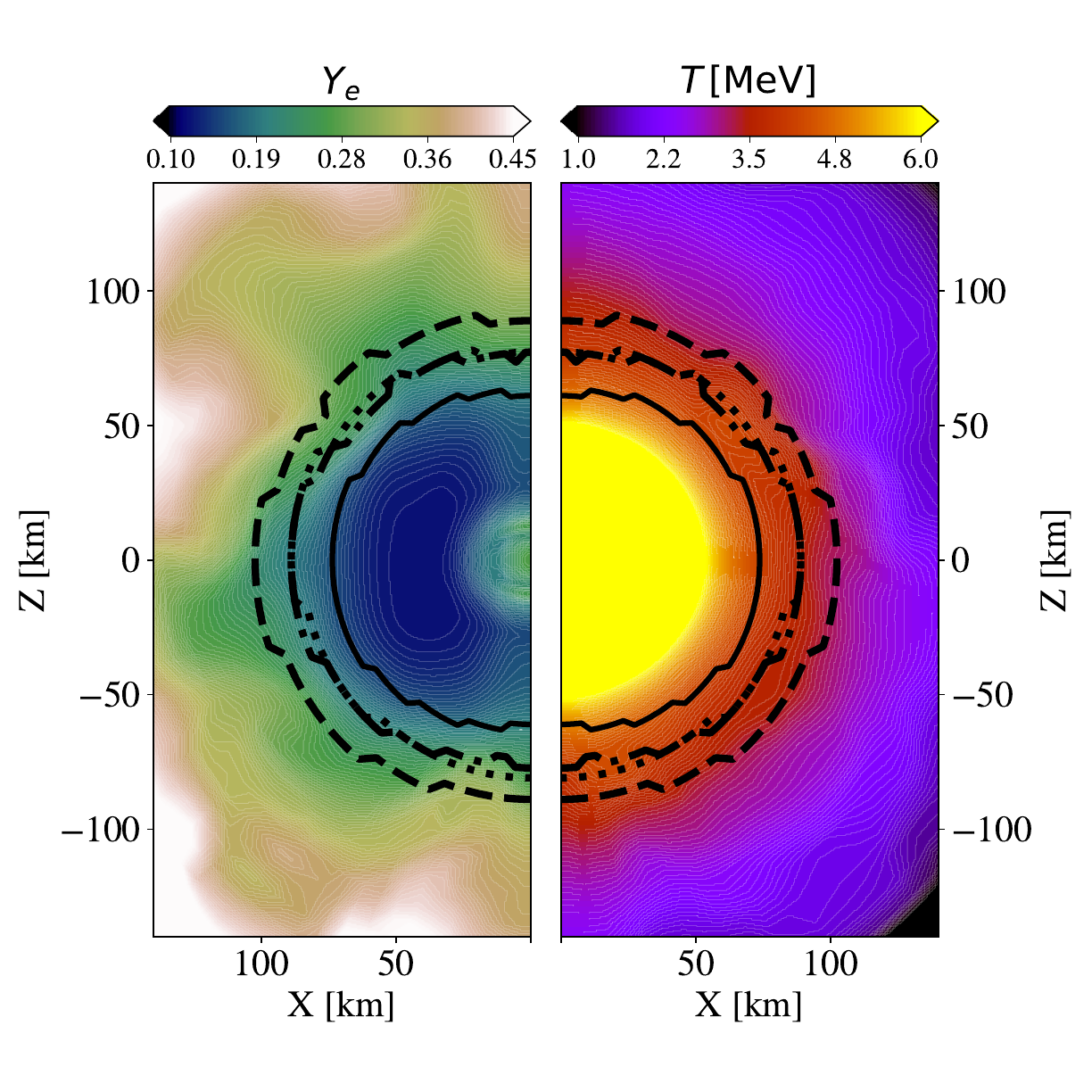}
    \includegraphics[width=0.5\textwidth,clip,trim={0 0 0 3.4cm}]{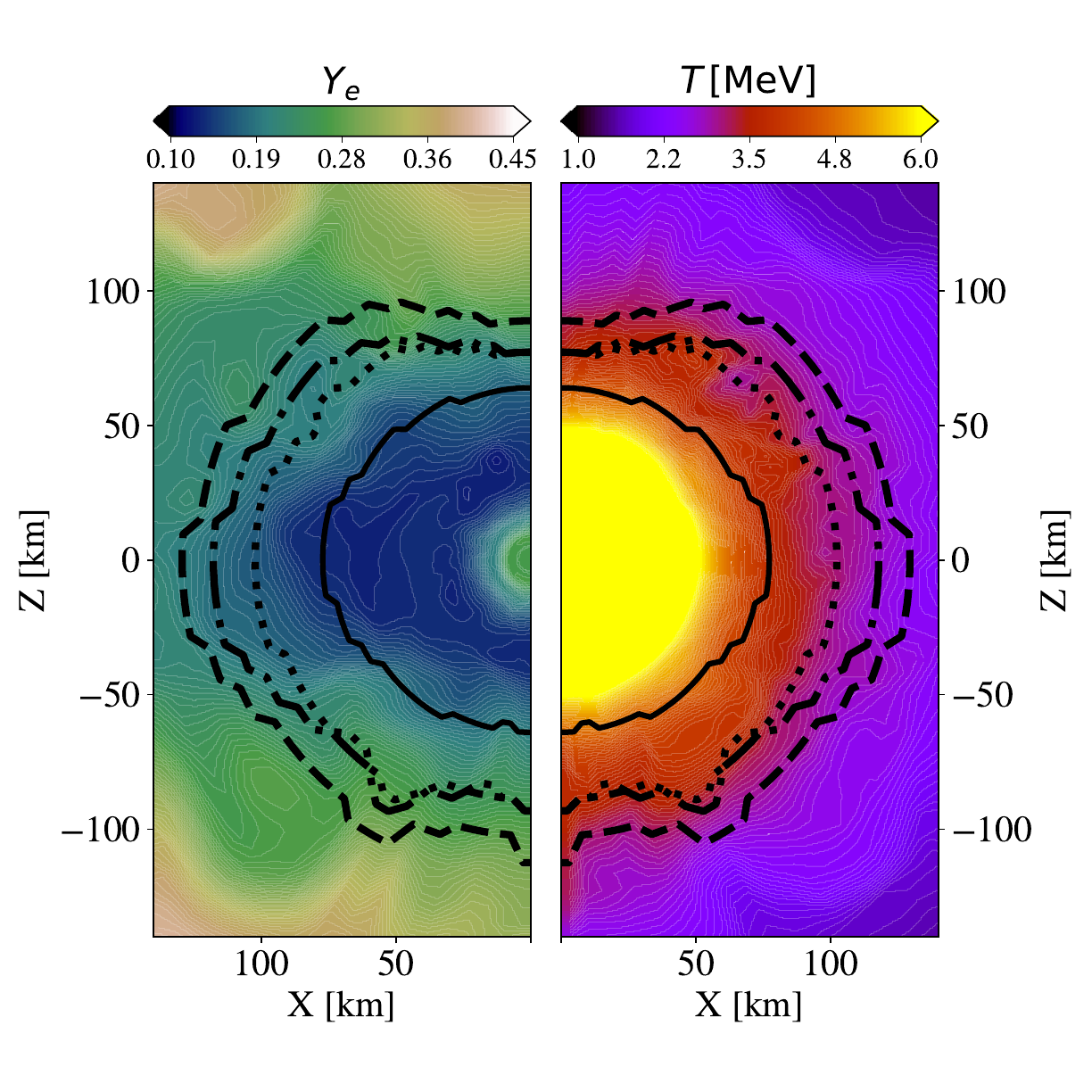}
    \caption{Electron fraction $Y_e$ (left panels) and temperature meridional distributions (right) for model \hydro{} (top) and \quadA{} (bottom) at $t=150$ ms p.b.
    The solid black curve represents the density threshold of $10^{11}$ g/cm$^3$, while the neutrinospheres for $\nu_e$, $\bar{\nu}_e$ and $\nu_x$ correspond respectively to the dashed, dotted and dash-dotted lines.}
    \label{fig:ye_T}
\end{figure}

This scenario is corroborated by the temperature $T$ distributions for models \hydro{} and \quadB{} at $t=150$ ms p.b. shown in the right panels of \refig{fig:ye_T}.
While the spatial maps for $T$ are quite similar between the two simulations at the given time, the location of the neutrinosphers differ significantly.
All three neutrinospheres are further from the center in the magnetized model along the equatorial plane, which corresponds to lower temperatures and consequently lower luminosities.
This result is consistent with previous axisymmetric models \citep{bugli2020} showing the impact of magnetic fields on the neutrino emission from the equatorial regions of the PNS.
\refig{fig:ye_T} shows another interesting feature of magnetized models: the distance between the $\nu_e$ and $\bar{\nu}_e$ neutrinospheres is larger w.r.t. model \hydro{}, which determines the differences between $L_{\nu_e}$ and $L_{\bar{\nu}_e}$.

\begin{figure}
    \centering
    \includegraphics[width=0.5\textwidth]{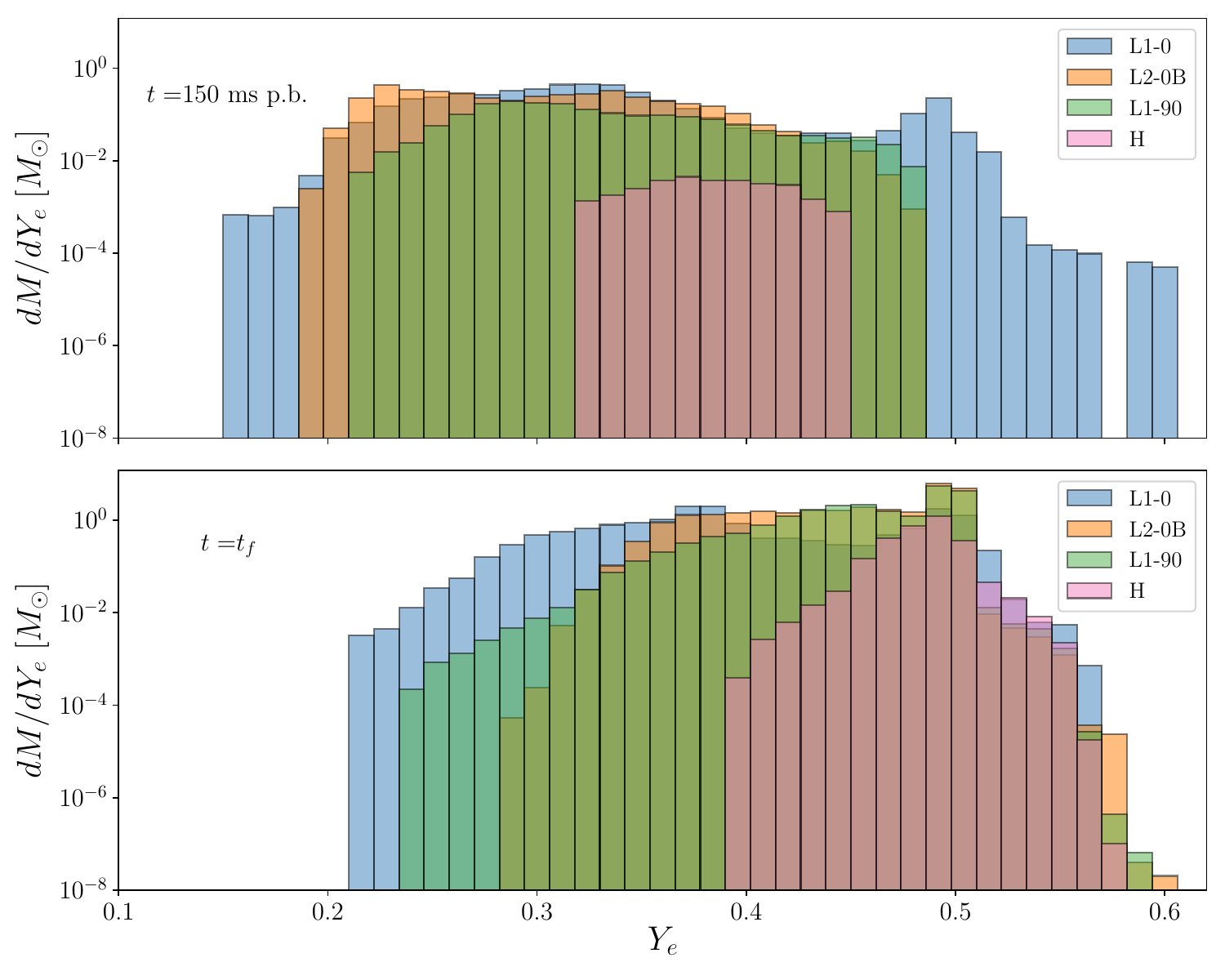}
    \caption{Ejecta mass fraction distribution over $Y_e$ at $t=150$ ms p.b. (top panel) and at the end of each simulation (bottom).}
    \label{fig:dMdYe}
\end{figure}

The origin of such phenomenon (which holds for all the magnetized models we produced) is connected to the $Y_e$ distribution among different simulations (left panels of \refig{fig:ye_T}).
Our magnetized models are characterized by a lower value of $Y_e$ in the explosion ejecta already 150 ms p.b. (top panel of \refig{fig:dMdYe}), since neutron-rich material is launched by the magnetorotational mechanism in the early post-bounce phase before it is stably accreted onto the PNS \citep{obergaulinger2021,reichert2021,reichert2022}.
Model \aligneddip{} has the highest mass in ejecta at $t=150$ ms p.b., as it is the only one producing a prompt explosion driven by its strong dipolar magnetic field.
It also shows tails of highly neutron-rich and proton-rich material in its early ejecta, while simulations with a quadrupolar and equatorial dipolar fields have early ejecta that cover a narrower range of $Y_e$.
However, all magnetized cases display quite broader distributions than the one seen for model \hydro{}, for which $Y_e$ ranges from $\sim0.32$ to $\sim0.45$ at $t=150$ ms p.b.
As time passes, the shape of all distributions changes significantly.
While the mass fraction of ejecta with $Y_e>0.5$ tends to reach roughly the same values regardless of the specific model, magnetized simulations systematically display more neutron-rich material than the hydrodynamic case.
Moreover, the distributions tend to shift towards higher values of $Y_e$, with model \quadB{} showing the most significant decrease of mass in neutron-rich ejecta.
This could either mean that some low $Y_e$ material previously unbound becomes once again bound to the PNS, or that the neutrino-matter interactions present a significant asymmetry between $\nu_e$ and $\bar{\nu}_e$. 

The immediate surroundings of the PNS tend to be clearly more neutron rich (right panels of \refig{fig:ye_T}), as the magnetic outwards transport of angular momentum provides extra support against the accretion of low $Y_e$ material onto the surface of the PNS.
This is consistent with model \quadB{} displaying the highest peak in low $Y_e$ material at $t=150$ ms p.b. (top panel of \refig{fig:dMdYe}) and the biggest decrease over time of neutron-rich material, since it has the most efficient transport of angular momentum in the equatorial region, where matter is not readily launched into the polar outflows.
As a result of a higher concentration of neutrons, an increased absorption of $\nu_e$ and emission of $\bar{\nu}_e$ leads to a larger gap between the two corresponding neutrinospheres, thus producing the observed differences in luminosity.
 
\section{Conclusions}
We conducted the analysis of hydrodynamic and magnetized 3d CCSN models with different initial magnetic field topologies, focusing on the development of the \LTWI{} and the associated multi-messenger emission.
Our simulations, which were first presented in \cite{bugli2021}, consider a massive and fast rotating stellar progenitor.
Although they all produce successful explosions, they show very different explosion dynamics and PNS evolution depending on the magnetic field configuration, which is reflected on the GW and neutrino emission during the accretion phase of the explosion.

The hydrodynamic model proves unstable to the \LTWI{}, which excites non-axisymmetric modes within the PNS with $l=1,2$ and large-scale, double-armed spiral structures after 200 ms p.b.  
Such deformations cause a strong burst in GW emission, which lasts several hundreds of ms with a frequency increasing steadily in time from $\sim$400 Hz to $\sim800$ Hz, spanning the sensitivity window of current terrestrial GW detectors.
Such results are consistent with the findings of recent numerical models disclosing the connection between the \LTWI{} and the multi-messenger emission of CCSN \citep{shibagaki2020,kuroda2020}.
Our magnetized models, on the other hand, show much weaker deformations within the PNS at a significantly lower frequency and with no clear growth of spiral arms during the duration of the simulations.
The onset of the \LTWI{} is quenched due to the outward magnetic transport of angular momentum, which depletes the regions close to the convective zone of differential rotation, thus stabilizing against the instability \citep{takiwaki2021}.
This results holds regardless of the specific topology of the magnetic field or evolution of the rotation profile, as long as the transport of angular momentum is efficient.
As a consequence, the corresponding emission of GW is significantly weaker w.r.t. the hydrodynamic case.
Moreover, the GW signal is characterized by a broader band spectral shape and it is not solely due to a large-scale oscillation mode of the PNS. 

In the hydrodynamic case the \LTWI{} leaves also a clear signature in the neutrino emission, in the form of oscillations with increasing frequency matching once again the oscillation modes of the PNS.
This confirms the recent analysis presented in \cite{shibagaki2021}, which highlights the strong correlation between the GW and neutrino signal modulated by the onset of the \LTWI{}.
Such correspondence is, however, not evident in the magnetized models, where the amplitude of the PNS oscillation modes is not strong enough to clearly let these spectral features rise above the broad-band signal generated at lower frequencies. 
However, all models present similar low-frequency signatures (at $F\lesssim$100 Hz) that are associated to non-axisymmetric oscillation modes of the shock front, i.e. SASI modes.
The only exception is the simulation with a strong aligned dipolar field, which produces a prompt magnetorotational explosion which proves too fast to allow for the growth of such modes.

Another aspect that marks a clear dichotomy between hydrodynamic and magnetized models is the lower neutrino luminosities along the equatorial plane produced in the latter (since the transport of angular momentum leads to more oblate PNS and inflates the neutrinospheres to colder regions), an effect significant enough to be potentially detectable by the KM3NeT neutrino observatory \citep{bendahman2022}. 
Moreover, the outward transport of angular momentum provides an extra rotational support to the surrounding neutron-rich in-falling matter, which lowers the $Y_e$ in the regions around the neutrinospheres and leads to a further distancing between the $\nu_e$ and $\bar{\nu}_e$ neutrinospheres.
Ultimately, this effect produces a clear difference in the corresponding luminosities $L_{\nu_e}$ and $L_{\bar{\nu}_e}$ which is not observed in the hydrodynamic case.
In addition to that, the ejecta composition tends to be systematically richer in neutrons, which can have critical consequences on the yields of r-process elements produced during the explosive nucleosynthesis \citep{reichert2021,reichert2022}. 

Our results show an overall dichotomy between the hydrodynamic model, which is dominated by the onset of the \LTWI{}, and the other models with strong magnetic fields of different topology.
The efficient magnetic transport of angular momentum tends to quench the growth of large-scale, non-axisymmetric modes within the PNS, suppressing their multi-messenger signatures.
On the other hand, spiral SASI modes can still develop significantly in those magnetized simulations that do not produce a prompt explosion.
This is due to the fact that the required advective-acoustic cycle is not impeded by the flattening of the rotation curve within the PNS and that magnetic fields do not necessarily stabilize SASI \citep{guilet2010}.
Moreover, all our magnetized models include strong rotation, which has been shown to favor the development of non-axisymmetric SASI modes \citep{yamasaki2008,kazeroni2017}.  
Since our models started with an already strong magnetic field, it remains unclear whether self-consistently including  the dynamo processes inside the PNS would lead to similar results.
Ultimately, it would depend on both the time-scale required to sufficiently amplify the magnetic fields and their spatial location, as the edge of the convective zone seems to play an important role in the onset of the \LTWI{}.


\section*{Acknowledgements}

he authors would like to thank Gianluigi Bodo, Rapha\"el Raynaud, and Paul Barr\`ere for stimulating discussions, as well as the anonymous referee for their useful suggestions.
MB and JG acknowledge support from the European Research Council (ERC starting grant no. 715368 -- MagBURST) and from the \emph{Tr\`es Grand Centre de calcul du CEA} (TGCC) and GENCI for providing computational time on the machines IRENE and OCCIGEN (allocation A0050410317).
MO acknowledges the support through the grant PID2021-127495NB-I00 funded by MCIN/AEI/10.13039/501100011033 and by the European Union, and the Astrophysics and High Energy Physics programme of the Generalitat Valenciana ASFAE/2022/026 funded by MCIN and the European Union NextGenerationEU (PRTR-C17.I1), from the the Deutsche Forschungsgemeinschaft (DFG, German Research Foundation) – Projektnummer 279384907 – SFB 1245, from the Spanish Ministry of Science via the Ramón y Cajal programme (RYC2018-024938-I), and from the the Generalitat Valenciana (grant PROMETEU/2019/071).
JG acknowledges the support from the PHAROS COST Action CA16214.

\section*{Data availability}
The data underlying this article will be shared on reasonable request to the corresponding author.




\bibliographystyle{mnras}
\bibliography{references} 




\bsp	
\label{lastpage}
\end{document}